\documentclass[useAMS,usenatbib]{mn2e}
\usepackage{graphicx}

\def\lesssim{\,\lower2truept\hbox{${<\atop\hbox{\raise4truept\hbox{$\sim$}}}$}\,}
\def\gtrsim{\,\lower2truept\hbox{${>\atop\hbox{\raise4truept\hbox{$\sim$}}}$}\,}

\title[Connecting the infrared and the X-ray background]
{Connecting the cosmic infrared background to the X-ray
background}
\author[Silva et al.]{L. Silva$^{1}$, R. Maiolino$^{2}$ \& G.L. Granato$^{3}$ \\
$^{1}$INAF - Osservatorio Astronomico di Trieste, Via Tiepolo 11, I-34131 Trieste, Italy\\
$^{2}$INAF - Osservatorio Astrofisico di Arcetri, Largo E. Fermi 5, I-50125 Firenze, Italy\\
$^{3}$INAF - Osservatorio Astronomico di Padova, Vicolo dell'Osservatorio 5, I-35122 Padova, Italy}

\begin{document}

\date{Accepted. Received}

\pagerange{\pageref{firstpage}--\pageref{lastpage}} \pubyear{2004}

\maketitle

\label{firstpage}

\begin{abstract}
We estimate the contribution of AGNs and of their host galaxies to
the infrared background. We use the luminosity function and
evolution of AGNs recently determined by the hard X-ray surveys,
and new Spectral Energy Distributions connecting the X-ray and the
infrared emission, divided in intervals of absorption. These two
ingredients allow us to determine the contribution of AGNs to the
infrared background by using mostly observed quantities, with only
minor assumptions. We obtain that AGN emission contributes little
to the infrared background ($<$5\% over most of the infrared
bands), implying that the latter is dominated by star formation.
However, AGN host galaxies may contribute significantly to the
infrared background, and more specifically 10--20\% in the
1--20$\mu$m range and $\sim$5\% at $\lambda<60\mu m$. We also give
the contribution of AGNs and of their host galaxies to the source
number counts in various infrared bands, focusing on those which
will be observed with Spitzer. We also report a significant
discrepancy between the expected contribution of AGN hosts to the
submm background and bright submm number counts with the
observational constraints. We discuss the causes and implications
of this discrepancy and the possible effects on the Spitzer far-IR
bands.
\end{abstract}


\section{Introduction}

Active Galactic Nuclei (AGN) are known to be the main contributors
to the cosmic X-ray background. Synthesis models suggested that
the shape of the X-ray background requires a mixture of unabsorbed
and absorbed AGNs (Setti \& Woltjer 1989; Comastri et al. 1995;
Gilli, Salvati, \& Hasinger 2001). This scenario has been verified
by the recent Chandra and XMM surveys which have resolved most of
the X-ray background up to energies of 10~keV (Brandt et al. 2001;
Giacconi et al. 2001; Hasinger et al. 2001). Indeed, the optical
identification of the sources making the X-ray background, as well
as their X-ray spectra, revealed both obscured and unobscured
AGNs, though unusual AGNs were also identified (e.g. Mainieri et
al. 2002; Barger et al. 2003; Fiore et al. 2003). The X-ray
surveys have also provided evidence that the evolution of AGNs is
different with respect to that known from previous optical
surveys. In particular, the evolution of low-luminosity AGN (in
the Seyfert range) peaks at much lower redshift (z$<$0.5) than for
more luminous, QSO-like AGNs (Hasinger 2003; Fiore et al. 2003;
Ueda et al. 2003). This luminosity dependent density evolution
(LDDE) seems to apply both to unobscured and obscured AGNs.

The circumnuclear gas responsible for the X-ray absorption is
associated with dust which absorbs the optical-UV radiation from
the AGN and reprocesses it thermally into the infrared. Dust
infrared emission has been observed in all AGNs (except in some BL
Lacs and radio-loud quasars where the IR radiation may be
dominated by the relativistically boosted synchrotron radiation).
At least in the near and mid-IR, it has been recognized that a
significant fraction of the dust radiation is due to heating by
the AGN (e.g. Granato, Danese, \& Franceschini 1997; Maiolino et
al. 1998a; Oliva et al. 1999; Glass 2004; Minezaki et al. 2004).
For this reason it has been argued that AGNs may contribute
significantly also to the IR cosmic background. This issue is of
utmost importance, since it would also have implications on the
fraction of IR background which should be ascribed to star
formation.

Some previous studies have estimated the contribution of AGN to
the IR background or to the source number counts (Granato et al.
1997; Matute et al. 2002; Risaliti, Elvis, \& Gilli 2002;
Andreani, Spinoglio, \& Malkan 2003).

The main ingredients for this estimate are the luminosity function
(LF) and evolution of AGNs and their spectral energy distribution
(SED). Previous works have adopted the observationally available,
still largely incomplete, LF of AGNs, that only very recently has
been better defined through Chandra and XMM surveys (Ueda et al.
2003). Generally, a single template for the AGN SED has been
assumed.

In this paper we investigate the contribution of AGNs to the IR
background and counts by using the most recent luminosity
functions and evolution obtained by the recent hard X-ray surveys.
We also use IR spectral energy distributions accurately derived by
means of nuclear IR data for a sample of Seyfert galaxies (both
obscured and unobscured) and which allow us to disentangle the AGN
IR emission from the contribution of the host galaxy. These data
also allow us to directly connect the IR emission of AGNs to their
intrinsic X-ray emission. Moreover, we use the novel approach of
dividing the observed IR SEDs into intervals of different AGN
obscuration, in terms of N$_H$ along the line of sight. This is
the same method adopted in the synthesis of the X-ray background.
These various pieces of information combined together allow us to
{\it directly connect the X-ray background (dominated by AGNs) to
the IR background without strong assumptions, by using only
observed and measured quantities.}

Throughout the paper we adopt the following cosmological
parameters: $\rm H_0 =70~km~s^{-1}~Mpc^{-1}$, $\rm \Omega _m =
0.3$ and $\rm \Omega _{\Lambda} = 0.7$.

\section{The luminosity function and evolution of AGN}

The most recent results from the hard X-ray surveys (both shallow
and deep) have been gathered by Ueda et al. (2003) to produce the
hard X-ray luminosity function and evolution of AGNs. These
luminosity functions derived in the hard X-rays are optimal for
our purposes, because they are much less affected by obscuration
than the luminosity function derived at other wavelengths.
Residual incompleteness in the fraction of heavily obscured AGN
has been accounted for (see discussion below for the case of
Compton thick AGNs). Ueda et al. (2003) derive these quantities
also by consistently requiring that the whole X-ray background is
properly fitted. They also match the optical luminosity functions
and evolution (Boyle et al. 2000) by assuming appropriate ratios
between X-ray and optical luminosity.

We direct the reader to Ueda et al. (2003) for more details on
their method and on the properties of their derived luminosity
functions. Here we only summarize two important results of their
work which are relevant for this paper. The density evolution of
AGNs is strongly dependent on their luminosity: the evolution of
Seyfert galaxies peaks at a much lower redshift (z$\sim$0.3--0.5)
than the more luminous QSOs (whose evolution peaks beyond z$>$1).
The other important result is that the various observational
constraints indicate that the fraction of obscured AGNs decreases
with luminosity. In particular, at high, QSO-like luminosities,
the fraction of obscured AGN is lower by a factor of about 2 with
respect to Seyfert nuclei.

In this paper we adopt the prescriptions by Ueda et al. (2003) for
the luminosity functions, evolution and distribution in absorbing
column densities $\rm N_H$, which properly reproduce the results
from the X-ray surveys and the X-ray background. For Compton thick
sources with $\rm 10^{24}<N_H<10^{25}cm^{-2}$, which contribute
significantly to the 30~keV bump of the X-ray background, but
which are poorly sampled by the current surveys at energies lower
than 10~keV, we adopt the same strategy of Ueda et al. by assuming
that their fraction is the same as for AGNs with $\rm
10^{23}<N_H<10^{24}cm^{-2}$ (consistent with the observations of
local AGNs, Risaliti, Maiolino, \& Salvati 1999).  Ueda et al.
(2003) also showed that this fraction of Compton thick sources
consistently reproduces the shape of the X-ray background
(especially at 30~keV).

Compton thick AGN with $\rm N_H>10^{25}cm^{-2}$ do not contribute
significantly to the X-ray background at any energy, since they
are totally absorbed and only their scattered component is
observed, and therefore they are generally neglected in the
synthesis models of the X-ray background. However, this class of
heavily absorbed AGN could contribute significantly to the IR
background and therefore this population must be included in our
model. We assume that the fraction of Compton thick AGN with $\rm
N_H>10^{25}cm^{-2}$ is the same as those with $\rm
10^{24}<N_H<10^{25}cm^{-2}$ and with $\rm
10^{23}<N_H<10^{24}cm^{-2}$, again consistent with the
observations of local AGNs (Risaliti et al. 1999).

\section{The nuclear infrared spectral energy distribution}

Various authors have derived the spectral energy distribution of
AGN in the IR (e.g. Elvis et al. 1994; Spinoglio, Andreani, \&
Malkan 2002; Alonso-Herrero et al. 2003; Kuraszkiewicz et al.
2003). However, in many cases their approach is not adequate for
our purposes.

At least for what concerns Seyfert galaxies we need to know the
nuclear infrared spectral energy distribution i.e. the infrared
light associated with the dust heated by the AGN and also direct
AGN emission (at this stage without the contribution by the host
galaxy). The infrared SED must be known both for unobscured AGNs
and for AGNs obscured by various levels of obscuration, and even
for the most obscured Compton thick AGN. We need a large
wavelength coverage, $\sim 1$ to 1000 $\mu$m in order to estimate
the contribution of AGNs to the near-IR to sub-mm background as
well as to the source counts in the Spitzer, ISO and SCUBA bands.
For each IR SED we also need to know its normalization in terms of
{\it intrinsic} hard X-ray flux, this will allow us to link the IR
SEDs to the hard X-ray luminosity functions.

\subsection{Spectral energy distribution of Seyfert nuclei}
\label{sec:sedseynuc}

We selected from the literature a large sample of Seyfert galaxies
for which {\it nuclear} near-IR and mid-IR observations have
detected clear signatures of {\it non-stellar} nuclear emission,
and where the stellar contribution has been properly removed.
Among these we only choose those objects having at least four
nuclear photometric points\footnote{As mentioned above,
by ``nuclear photometry'' we mean the central non-stellar flux.
 In many works this
flux is extracted through a spatial decomposition of the
near/mid--IR images into a stellar light profile and a nuclear
point like source, or even spectroscopically (through the dilution
of the stellar features, see reference in the Appendix). In cases
where mid-IR images are not available (and only fixed aperture
fluxes are available), if from the K-band image the nuclear source
is found to dominate the light within the central few arcseconds,
then it is assumed that also the mid-IR flux dominates within the
same aperture (this is justified by the fact that the contrast
between AGN hot-dust emission and stellar light is higher in the
mid-IR).} and at least one of them must be in the mid-IR (10$\mu$m
and/or 20$\mu$m); such a minimum set of data is required for a
proper interpolation with the models discussed below. We also
required that the sources had hard X-ray measurements good enough
to provide the intrinsic (i.e. unabsorbed) hard X-ray flux and a
measure of the absorbing column density $\rm N_H$. In particular
we require that the X-ray data allow a measurement of N$_H$ with
an accuracy of about 50\% at least, or that they allow to set an
upper limit on N$_H$ below 10$^{22}$~cm$^{-2}$ (below this value
the absorption of the 2--10~keV luminosity is negligible). A total
of 33 Seyfert nuclei matched these requirements (see Appendix).
Most of these Seyferts are from the Maiolino \& Rieke (1995)
sample. The list of Seyferts used to derive the IR SEDs, along
with the references for the data and some additional details are
given in the Appendix.

The nuclear IR data were then {\it interpolated} with the updated
models of Granato \& Danese (1994). These models were generated by
means of a detailed radiative transfer code for dust heated by a
nuclear central source with a typical AGN spectrum. The code
includes various possible geometries and physics for the
distribution of the circumnuclear dust (torus with variable
height, radii and density, as well as tapered disks). The code
also allows variation in the distribution of dust grain sizes, to
account for possible deviations from the standard ISM extinction
curve as suggested by some recent works (Maiolino, Marconi, \&
Oliva 2001; Maiolino et al. 2001; Gaskell et al. 2003). There are
several other models which have been proposed to describe the
circumnuclear dust emission in AGNs (e.g. Pier \& Krolik 1993;
Efstathiou \& Rowan-Robinson 1995; Nenkova, Ivezic, \& Elitzur
2002). However, we do not want to discuss here the different
features of the various models and which of them may be more
appropriate to describe the properties of the circumnuclear dust
in AGN. In this paper the models are simply used to interpolate
the observed nuclear IR data with a physically plausible SED. As
long as the latter is constrained by the data, the choice of the
specific model does not change the results concerning the
interpolated spectral region, typically within $\sim 2 - 20 \mu$m.

More critical is the {\it extrapolation} of the models beyond
20$\mu$m, where nuclear IR data are not available (only large
aperture data from ISO or IRAS are available at $\lambda >
20\mu$m). In this spectral region there is some degeneracy among
models. Our best fitting models (for each individual AGN of our
sample) generally suggest a drop of the IR emission redward of
$\sim$30-50$\mu$m. Some authors propose models which fit
relatively flat SED to data up to 100$\mu$m (Andreani,
Franceschini, \& Granato 1999; Kuraszkiewicz et al. 2003). As
discussed in several works on IR emission by AGNs, MIR data are
generally better reproduced by models in which the SED drops in
the FIR (e.g. Granato \& Danese 1994; Efstathiou \& Rowan-Robinson
1995; Galliano et al. 2003). Moreover those models with relatively
flat FIR SEDs would require the existence of very extended dusty
torii ($\sim 1-10$ kpc, e.g. Andreani et al. 1999). This is
difficult to justify, and in any case, it has been excluded at
least for NGC1068, whose torus has been directly observed (Jaffe
et al.
2004).\\
However, the far-IR part of our nuclear AGN SEDs remains
subject to strong uncertainties and admittedly model dependent.\\

   \begin{figure}
   \centering
   \includegraphics[angle=0,width=8cm]{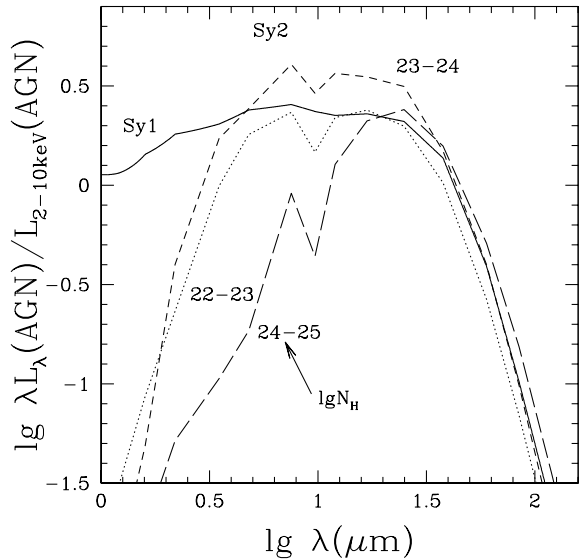}
  \caption{Nuclear (AGN only) infrared spectral energy distributions of Seyfert
  galaxies, normalized to the hard X-ray (2-10~keV)
  intrinsic luminosity and averaged within
  bins of absorbing N$_H$. The solid line is the SED for Sy1s, the dotted line
  for Sy2s with $\rm 10^{22}<N_H<10^{23}cm^{-2}$, the short-dashed line for
  Sy2s with $\rm 10^{23}<N_H<10^{24}cm^{-2}$ and the long-dashed line is for
  Sy2s with $\rm 10^{24}<N_H<10^{25}cm^{-2}$.}
              \label{nuc_sed}%
    \end{figure}

The SEDs of the single objects were then normalized by the
intrinsic, unabsorbed X-ray flux in the 2-10~keV band. Note that
this prevents us using objects with $\rm N_H > 10^{25}cm^{-2}$
(such as NGC1068, Matt et al. 1997), because their X-ray radiation
it totally absorbed at all energies and therefore the intrinsic
X-ray emissions cannot be recovered. The issue of the inclusion of
this class of objects in our model will be discussed in Section 5.

The X-ray--normalized infrared SEDs were then divided into
intervals of absorption in terms of $\rm N_H$. We do not enter
into the issue of the mismatch between gas absorption in the
X-rays and dust absorption in the optical/infrared (Maiolino et
al. 2001) and, more specifically, that dust absorption is
generally significantly lower than expected by a Galactic
dust-to-gas ratio and by a Galactic extinction curve. Regardless
of such physical issues, we simply follow the phenomenological
approach of dividing the AGNs in terms of N$_H$; this will then
allow to properly match the {\it observationally} more adequate IR
SED to each class of absorbed AGN used to synthesize the X-ray
background. The resulting average infrared SEDs are shown in
Fig.~\ref{nuc_sed}. Note that in our sample of Sy2 there are no
objects with $\rm 10^{21}<N_H<10^{22}cm^{-2}$, the inclusion of
this class of AGN (which is only a minor fraction of the whole
population, as found both locally and in the hard X-ray surveys,
Risaliti et al. 1999, Ueda et al. 2003) will be discussed in
Section 5.

The main difference between the SEDs of Sy1 and the SED of Sy2s
with $\rm 10^{22}<N_H<10^{23}~cm^{-2}$ is absorption in the
near-IR at $\rm \lambda \la 2\mu m$ and some mild Silicate
absorption at 9.7$\mu$m. For Sy2s with $\rm
10^{23}<N_H<10^{24}~cm^{-2}$ the shape of the IR SED does not
change significantly except for a slight, overall increase of the
emission. The finding that the shape of the IR SED remains
unchanged indicates that the medium absorbing the X-rays with $\rm
N_H \sim 10^{23.5}~cm^{-2}$ is either too small in size to obscure
the near-IR emitting region or that its dust content/composition
is unable to effectively absorb the near-IR radiation. The slight
increase of the overall IR emission might indicate a slight
increase of the covering factor of the circumnuclear dust. However
one should keep in mind that the number of objects used to create
the average SEDs is still small and therefore statistical
fluctuations may still be significant. For Compton thick Sy2s
($\rm N_H>10^{24}~cm^{-2}$) the IR SED is significantly different,
showing a prominent absorption even in the mid-IR (although still
much lower than expected by the dust associated with a Compton
thick medium and with a Galactic gas-to-dust ratio and
composition).

\subsection{Spectral energy distribution of quasars}
\label{sec:sedqsonuc}

Disentangling the nuclear infrared emission from the contribution
of the host galaxy is a much more difficult task for quasars.
Indeed at their larger distances, even small apertures include a
significant fraction of the host galaxy. Resolving the host
galaxies in several quasars has been achieved by various authors
in the near-IR (McLeod \& Rieke 1994a,b; Surace, Sanders, \& Evans
2001), but at longer wavelengths ($\lambda \ge 3 \mu$m) we still
have integrated information. Moreover, infrared data are generally
available for unobscured quasars, while there is little
information on the IR emission of the few type 2 quasars known.
For these reasons we assume that the shape of the infrared
emission as a function of the obscuration is the same as for the
Seyfert nuclei.

However the normalization to the X-ray flux cannot be assumed to
be the same, since it is known that the X-ray radiation decreases
relative to (at least) the optical-UV radiation at high
luminosities. To determine the normalization factor we adopt again
an observational approach. We merged the samples of quasars
measured in the infrared by Andreani et al. (1999, 2003) and Haas
et al. (2000, 2003). These samples are all selected in the optical
and, therefore, include only unobscured quasars. We selected only
those quasars having at least four IR detections, at least one of
which is in the far-IR, thus allowing a proper determination of
their total IR SED. The latter criterion may bias our sample in
favor of IR-bright quasars and with excess of star formation, as
discussed below. We further selected only those quasars which also
have X-ray measurements (good enough to ensure that there is no
significant N$_H$ absorption affecting the observed
L$_{2-10keV}$). Then we assume that at 12$\mu$m the QSO integrated
radiation is dominated by the QSO light, which is a reasonable
approximation, since at this wavelength the AGN radiation is
strong while the radiation from the galaxy is close to its
minimum. We determine the average ratio $\rm \lambda
F_{\lambda}(12\mu m)/F(2-10keV)$ for all quasars in our sample and
use this quantity to scale the SED derived for Sy1s relative to
the X-ray emission. Then the SED of obscured quasars are derived
from the SED of Sy2s by re-scaling them to maintain the same
intensity relative to unobscured quasars (i.e. the same relative
intensities shown in Fig.~\ref{nuc_sed}). The average ratio
$\lambda F_{\lambda} (12\mu m)/F(2-10keV) = 3.6$ derived for
quasars is only a factor 1.4 higher than for Sy1s (2.5). This
result may sound odd, since when looking at the optical-UV bands
the ratio F(opt)/F(X) is higher by a factor of three in quasars
relative to Seyferts. Our finding is suggestive that the covering
factor of dust is lower at higher luminosities, which reduces the
overall IR flux relative to the optical radiation. This issue has
been briefly discussed in Maiolino (2002) and will be subject of a
forthcoming more detailed paper (Maiolino \& Granato, in prep.).
This finding is also in agreement with the recent results by Ueda
et al. (2003) and by Hasinger (2003) who find that the fraction of
obscured AGN at high, quasar-like luminosities is lower than in
Seyferts.

\section{The total infrared spectral energy distribution of AGNs and of their
host galaxies}

We have also investigated the contribution by the host galaxies of
AGN. This is a more uncertain investigation since it relies on the
assumption that the relation between host galaxy and nuclear
emission is, at high redshift, the same as observed in local
galaxies.

The additional problem is that our method requires a
proportionality between infrared SED and X-ray luminosity. There
is no reason, in principle, to assume that the host galaxy
infrared light is proportional to the AGN-dominated X-ray light
(this is a conceptual problem also of other works, as discussed in
Section~7). However, various studies have shown that the two
phenomena, star formation and AGN, are related and must yield to
the well known correlation between Black Hole mass and bulge mass
(Ferrarese \& Merritt 2000; Marconi \& Hunt 2003). In particular,
Croom et al. (2002) found that the {\it optical} luminosity of the
host galaxy scales as $\rm L_{host}\propto L_{QSO}^{0.4}$,
although with a large scatter. We therefore investigated the SEDs
of the host galaxies as a function of the (X-ray) luminosity of
the active nucleus.

We collected all infrared measurements which, especially for
Seyferts, include the contribution of both the nucleus and the
host galaxy (e.g. IRAS and 2MASS total photometry). We then
subtracted from the integrated photometric points the contribution
of the nucleus estimated as discussed in Section~3. The residuals
were then fitted with galaxy models (a combination of active and
quiescent stellar populations) obtained with the publicly
available code GRASIL\footnote{The executable and a set of galaxy
SEDs including those presented in Silva et al. (1998) can be
downloaded at {\it
http://adlibitum.oat.ts.astro.it/silva/default.html} or {\it
http://web.pd.astro.it/granato/}} (Silva et al. 1998).

The resulting SEDs of the host galaxies were normalized by the
X-ray luminosity of the nucleus (which is required to link
the SEDs to the AGNs evolution and luminosity functions), grouped
in intervals of luminosity and then averaged within each group.

\begin{figure}
   \centering
   \includegraphics[angle=0,width=8cm]{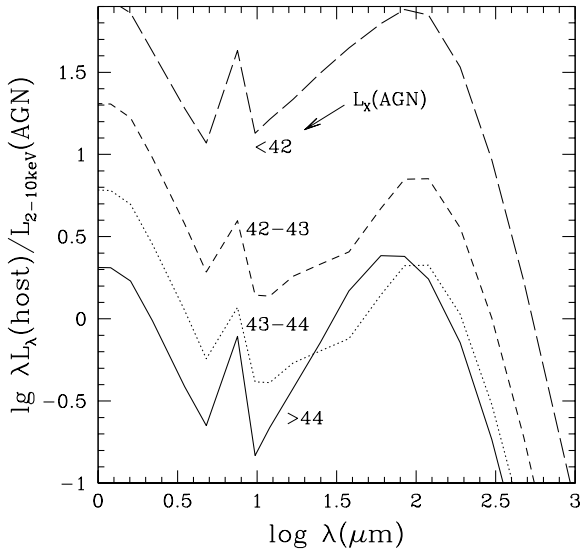}
  \caption{Infrared SED of Seyfert's host galaxies normalized to the
  X-ray luminosity of the AGNs. The SED have been divided in bins of
  X-ray luminosity and averaged. The solid line is the SED of galaxies
  hosting AGNs with $\rm L_{2-10keV}>10^{44}erg~s^{-1}$, dotted for
  galaxies hosting AGN with $\rm 10^{43}< L_{2-10keV}<10^{44}erg~s^{-1}$,
  short-dashed for galaxies hosting AGN with
  $\rm 10^{42}< L_{2-10keV}<10^{43}erg~s^{-1}$ and long-dashed for
  galaxies hosting AGN with $\rm L_{2-10keV}<10^{42}erg~s^{-1}$}
              \label{host_sed}%
    \end{figure}

Fig.~\ref{host_sed} shows the infrared SEDs of the host galaxies,
normalized to the X-ray luminosity, and averaged within various
ranges of L$_X$. As expected the host galaxy contribution relative
to the active nucleus decreases at higher AGN luminosities. At
quasar-like luminosities ($\rm L_X > 10^{44}erg~s^{-1}$) the
luminosity of the host decreases in the near-IR but not
significantly in the far-IR. This may indicate that the hosts of
high luminosity AGNs are undergoing to enhanced star formation,
similar to that found by the SDSS (Kauffmann et al. 2003), but may
also be a bias due to our selection criterion for quasars in our
sample (Section~3.2).

The total SEDs were simply obtained by combining the nuclear SED
(Section~3) and the SEDs from the host galaxies for each
range of luminosity ($\rm L_X$) and AGN absorption ($\rm N_H$).

\section{Connecting the infrared and X-ray background}

The IR SEDs divided into groups based on their level of absorption
and normalized to the hard X-ray flux, along with the hard X-ray
luminosity functions and evolution by Ueda et al. (2003), provide
us all the information needed to connect the X-ray cosmic
background to the infrared background.

For AGNs with (mild) absorption in the range $\rm
10^{21}<N_H<10^{22}cm^{-2}$ we have no IR SED (see Appendix),
since this is a rare class of AGN in the local universe and
therefore there are no representatives in our sample. However at
$\rm N_H<10^{22}cm^{-2}$, even for a Galactic dust-to-gas ratio,
the absorption in the IR is negligible (less than 0.2~mag even in
the 2$\mu$m  band). Therefore, for this class of AGN we assume the
same IR SED as type 1 Seyfert, and with the optical direct
component absorbed (yet the optical spectrum is not of interest
since it is not investigated in this paper).

Compton thick AGNs absorbed by $\rm N_H > 10^{25}cm^{-2}$ are not
included in Ueda et al. (2003) because their contribution to the
X-ray background is negligible (totally absorbed at all energies
and only a weak reflection component is observed in the X-rays).
However, their reprocessed radiation is observed in the infrared
(the stereotype case is NGC1068). As discussed in Section~3, we
cannot derive the IR SED of this class of objects normalized to
the X-ray emission just because it is not possible to measure
their intrinsic X-ray luminosity (and anyhow only NGC1068 has the
required data to derive a nuclear IR SED). Therefore we assume
that absorbed AGNs with $\rm N_H > 10^{25}cm^{-2}$ have the same
IR SED as the AGNs with $\rm 10^{25}> N_H > 10^{24}cm^{-2}$ both
in terms of shape and in terms of normalization relative to the
intrinsic X-ray emission. We further assume, based on the $\rm
N_H$ distribution of local Seyferts (Risaliti et al. 1999), that
the fraction of AGN with $\rm N_H > 10^{25}cm^{-2}$ is the same as
the fraction of AGNs with $\rm 10^{25}> N_H > 10^{24}cm^{-2}$.

Another issue is the definition of the dividing luminosity between
``Seyferts'' and ``quasars'', i.e. what are the luminosity ranges
where the two different classes of SEDs should be used. This issue
is actually little important since as discussed in
Section~\ref{sec:sedqsonuc} the nuclear SEDs for Seyferts and
quasars are nearly identical (just 40\% difference in the
normalization). Traditionally, in the X-rays the dividing
luminosity is taken at about $\rm 10^{44}erg~s^{-1}$ in the
2--10~keV band, which roughly corresponds to a bolometric
luminosity of $\rm 10^{45}erg~s^{-1}$. The luminosity of $\rm
L_{2-10keV}=10^{44}erg~s^{-1}$ appears to be an adequate dividing
luminosity also for the Seyfert 1s and the quasars in our samples
(see Appendix), and therefore we will consistently use this
dividing luminosity.

To compare the contribution of AGN to the IR bands with that due
to normal/starburst galaxies (i.e. galaxies powered only by
stars), we have also included the contribution by the latter, as
described in Silva et al. (2004). There, the galaxy populations
accounted for are spheroidal galaxies (based on the model by
Granato et al. 2004), and late-type galaxies (spirals and
starbursts). We refer to the cited papers for more details.

\section{Results}

\subsection{The contribution of AGNs to the infrared background}

\begin{figure*}
   \centering
   \includegraphics[angle=0,width=17truecm]{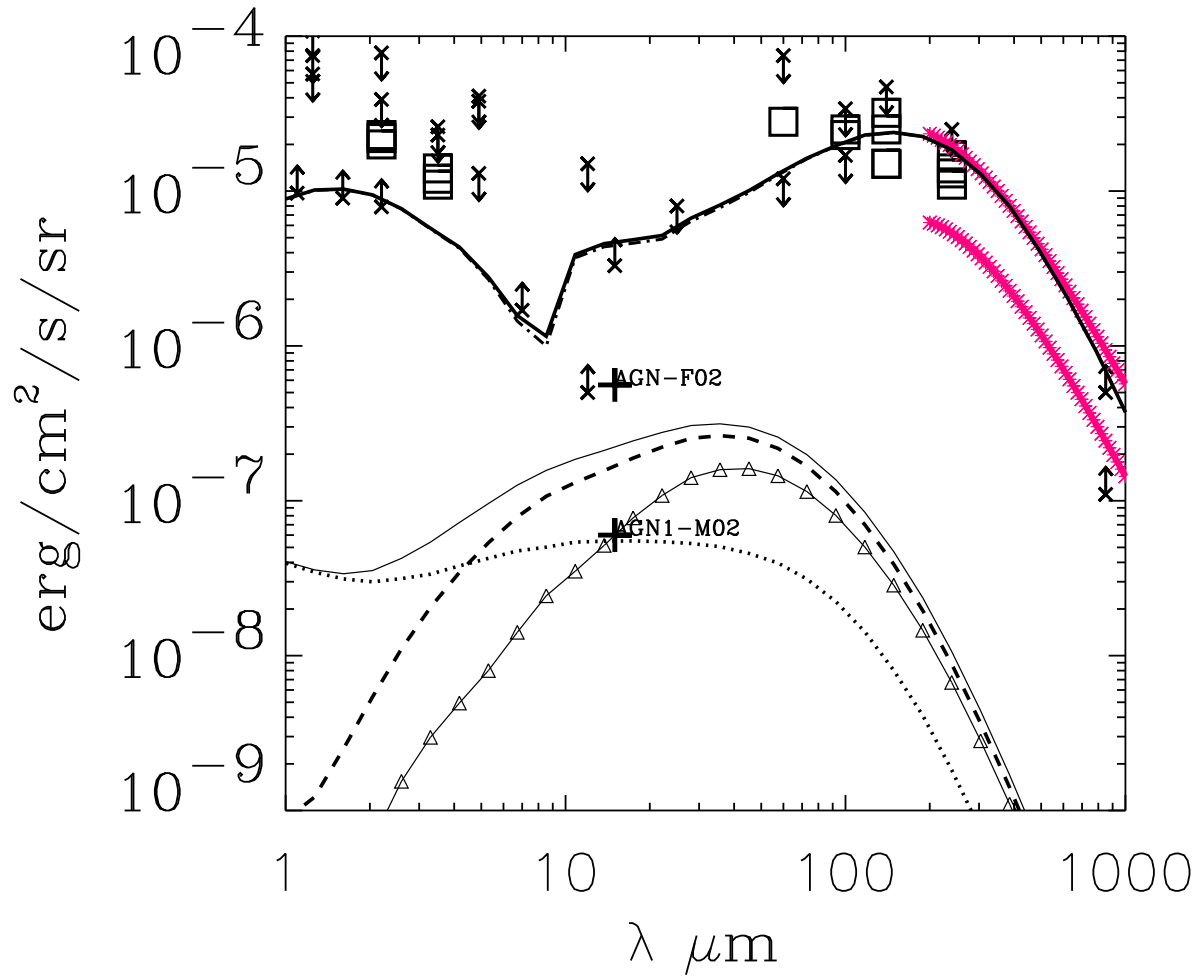}
   \caption{The IR background. The contribution due to the nuclear emission by AGN is shown by the thin continuous
   line. The dotted and dashed lines show the contribution by type 1 and 2 AGNs respectively.
   The connected triangles show the contribution by Compton
   thick nuclei (N$_H>10^{24}$ cm$^{-2}$). The IR background due to galaxies is shown by the dot-dashed
   line. Observational data for the IR background
   are from Hauser \& Dwek (2001). The thick crosses show the contribution to the $15 \mu$m
   background estimated for type 1 AGN by Matute et al. (2002),
   and for total AGN by Fadda et al. (2002).}
    \label{bg_nuc}
    \end{figure*}

Our estimated contribution of AGN (without hosts) to the IR
background is shown in Fig.~\ref{bg_nuc}.

   \begin{figure*}
   \centering
   \includegraphics[angle=0,width=17truecm]{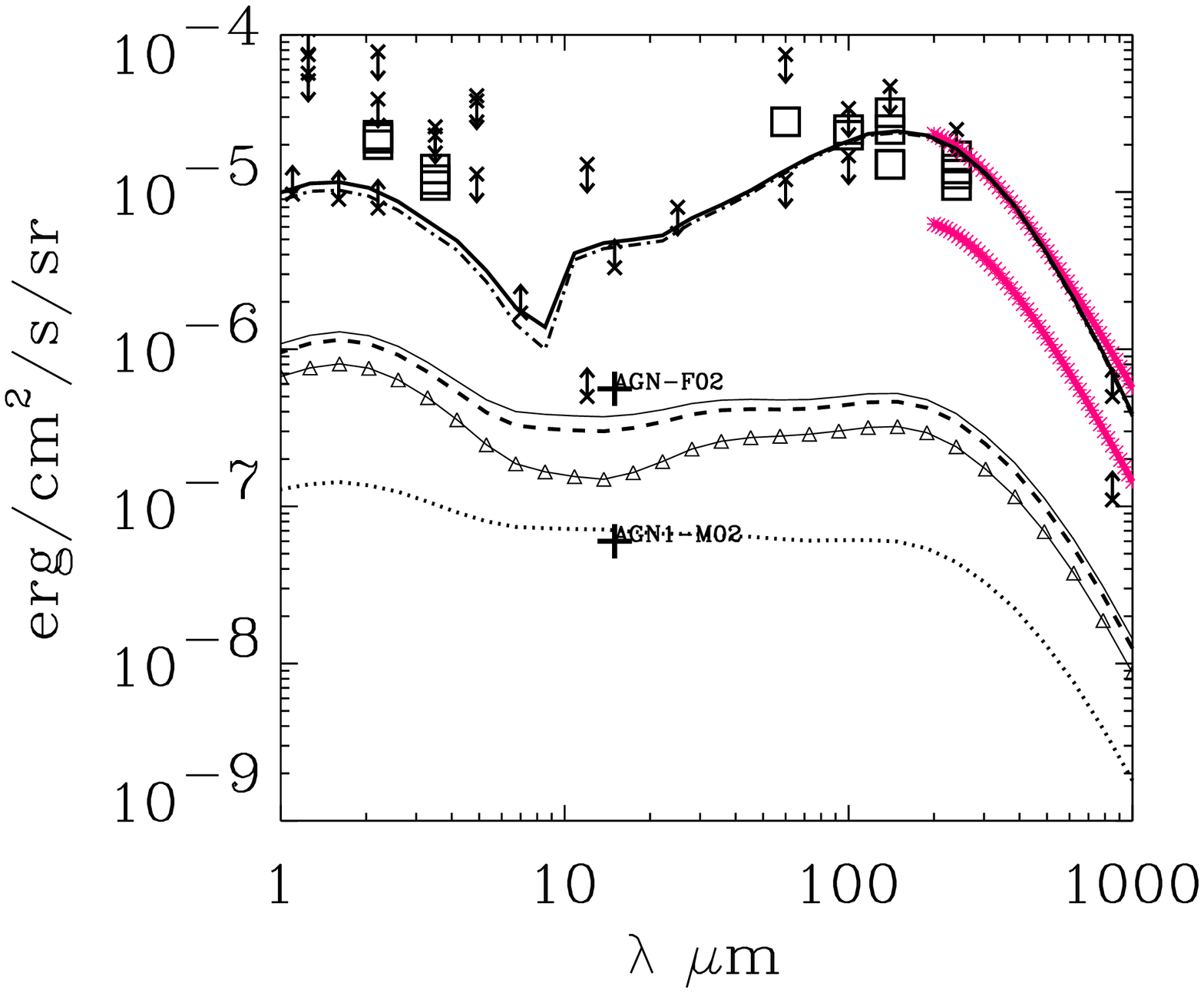}
   \caption{The IR background. The contribution due to the emission by the AGNs + their host galaxies
   is shown by the thin continuous line. The dotted and dashed lines show the contribution by
   type 1 and 2 AGNs+hosts respectively.
   The connected triangles show the contribution by Compton
   thick systems (N$_H>10^{24}$ cm$^{-2}$). The IR background due to galaxies,
   the same as in Fig. \ref{bg_nuc}, is shown by the dot-dashed
   line. Observational data for the IR background
   are from Hauser \& Dwek (2001). The thick crosses show the contribution to the $15 \mu$m
   background estimated for type 1 AGN by Matute et al. (2002), and for total AGN by Fadda et al. (2002).}
    \label{bg_ext}
    \end{figure*}

The main result is that AGN contribute little to the overall IR
background. The spectral region where some significant AGN
contribution is found (3 to $\sim$5\%) is between 5$\mu$m and
40$\mu$m. At wavelengths just below $10 \mu$m, the background due
to galaxies show a leap, that causes a maximum ($\ga 10$\%) for
the contribution of AGN. But that leap is mostly due to the
transition between the evolutionary regime assumed for late-type
galaxies in the MIR to sub-mm region and in the NIR (see Silva et
al.\ 2004)

At all infrared wavelengths $\lambda >5\mu$m the AGN contribution
is dominated by heavily obscured Seyfert nuclei ($\rm
N_H>10^{23}cm^{-2}$), in particular at $\lambda > 20\mu$m the
contribution is dominated by Compton thick Seyferts ($\rm
N_H>10^{24}cm^{-2}$).

In Fig.~\ref{bg_nuc} we also show the contribution of AGN to the
15$\mu$m background inferred by Fadda et al. (2002) and by Matute
et al. (2002) (limited to type 1 AGN only) based on the
cross-correlation between hard X-ray sources and mid-IR ISO
sources or through direct optical identification of the ISO
sources. These values are higher than predicted by our model
because the ISO sources also include the contribution by the host
galaxies. Therefore the comparison with these results will be
discussed in the next Section.

We point out that recently growing evidence has been found for a
class of totally buried and optically elusive AGN in nearby
galaxies (Della Ceca et al. 2002; Maiolino et al. 2003; Marconi et
al. 2000). This class of AGNs may be as numerous as classical,
optically identified AGN. Most of them are Compton thick and
therefore, if they are present also at high redshift, they would
mostly be missed even by current hard X-ray surveys. On the other
hand their IR emission would provide an additional contribution to
the IR background, which at the current stage is not included in
our model.

We also recall that, as described in Section 3, the far-IR nuclear
SEDs of AGN is not directly observed, therefore our estimated
background in that spectral region is subject to higher
uncertainties as compared to the near and mid-IR.

The contribution of AGNs to the IR source counts is discussed only
for the case of the AGN+host SEDs in the next Section, since
available data refer to the total emission of sources.

\subsection{The contribution of AGN {\it and} of their
host galaxies to the infrared background}
\label{sec:agnhost}

   \begin{figure*}
   \centering
   \includegraphics[angle=0,width=8.5truecm]{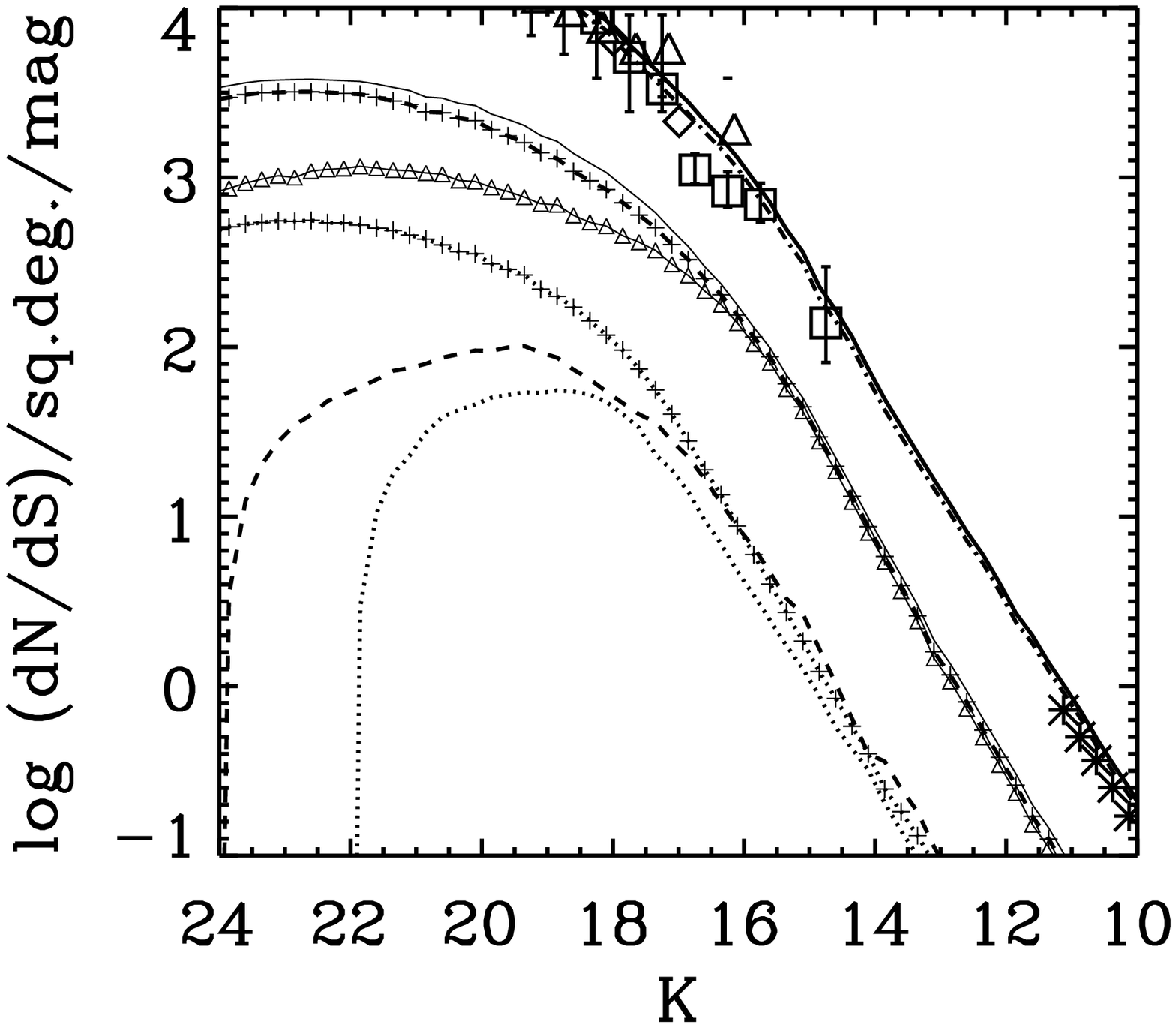}
   \includegraphics[angle=0,width=8.5truecm]{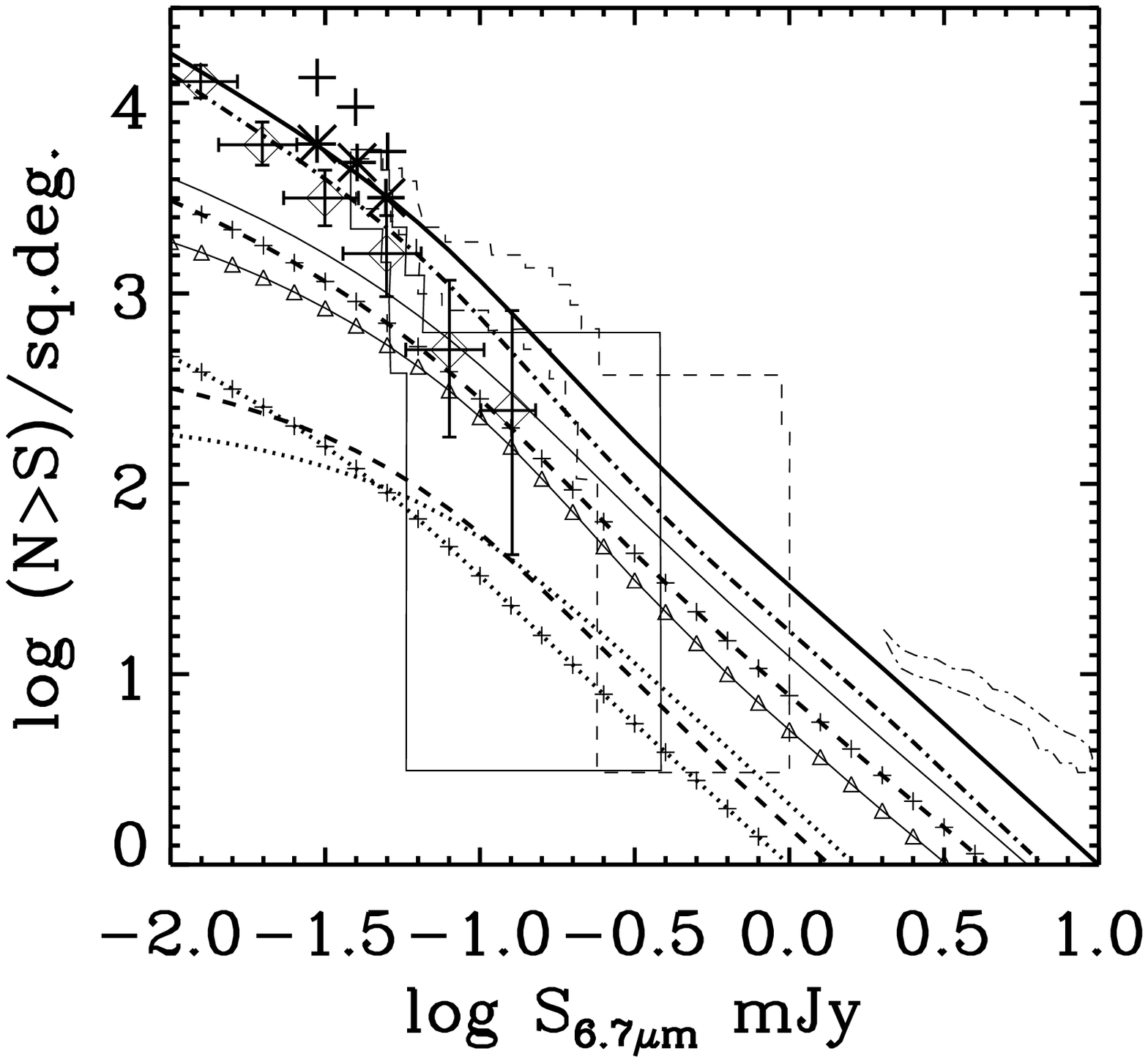}
   \includegraphics[angle=0,width=8.5truecm]{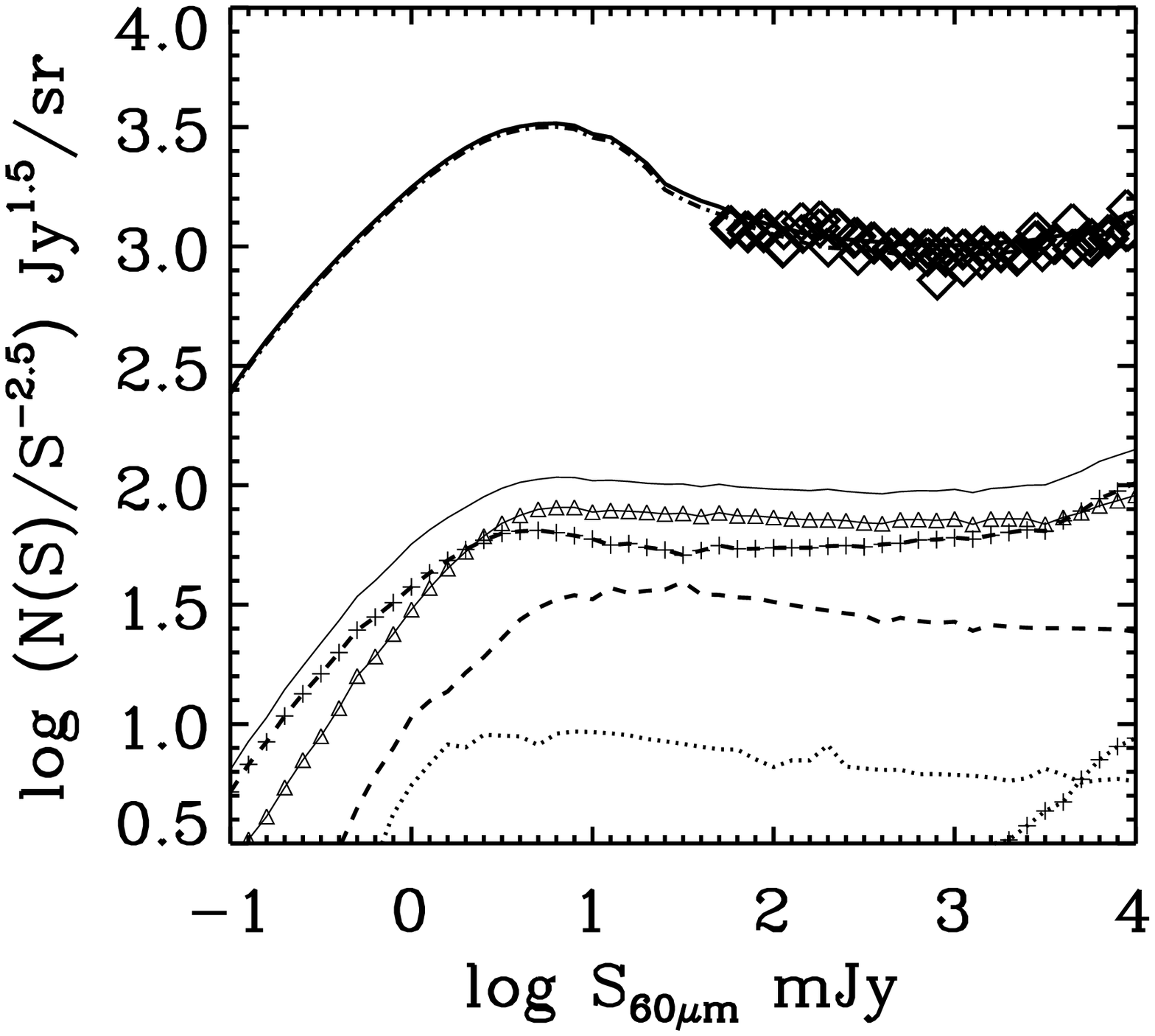}
   \includegraphics[angle=0,width=8.5truecm]{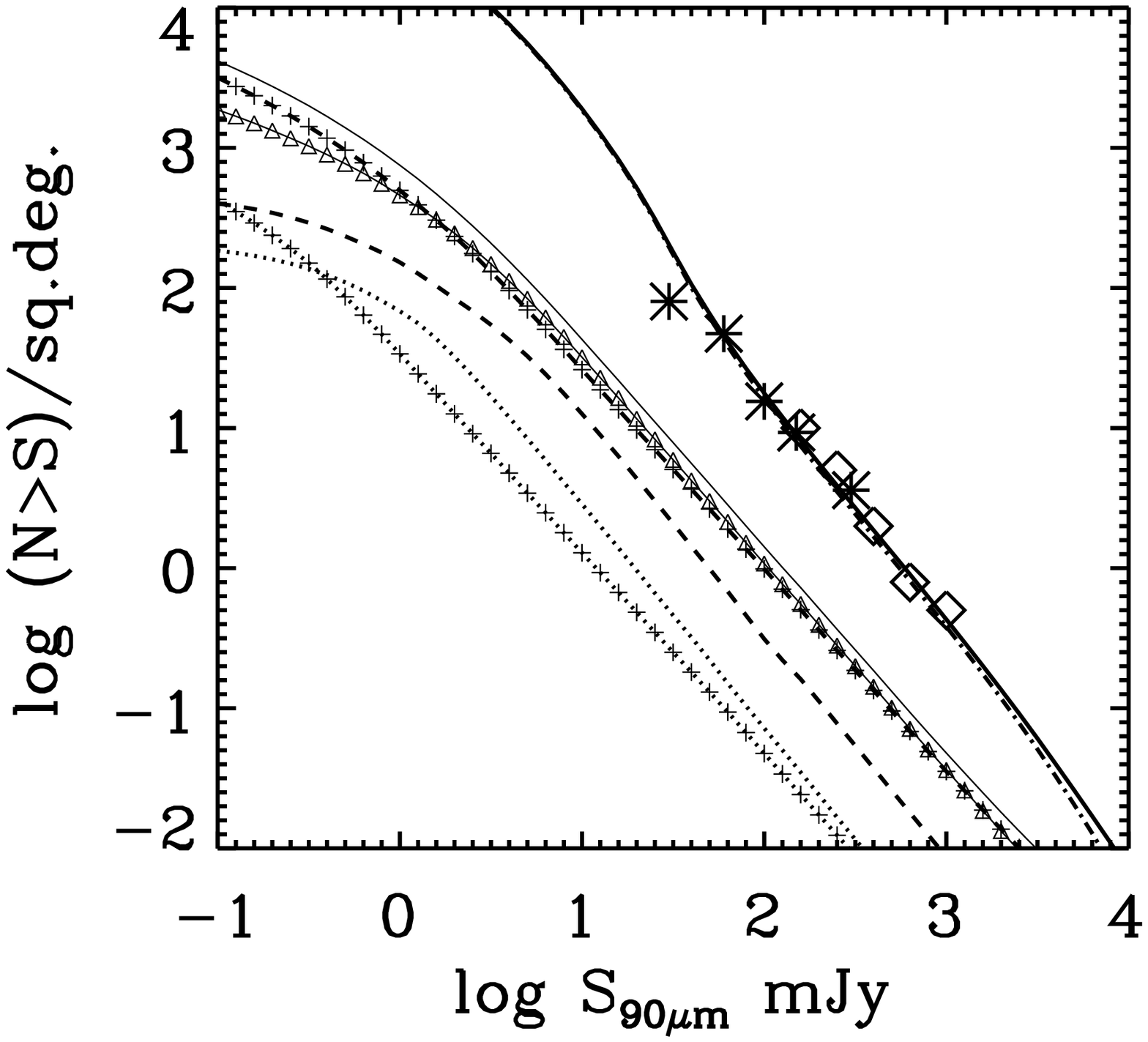}
   \includegraphics[angle=0,width=8.5truecm]{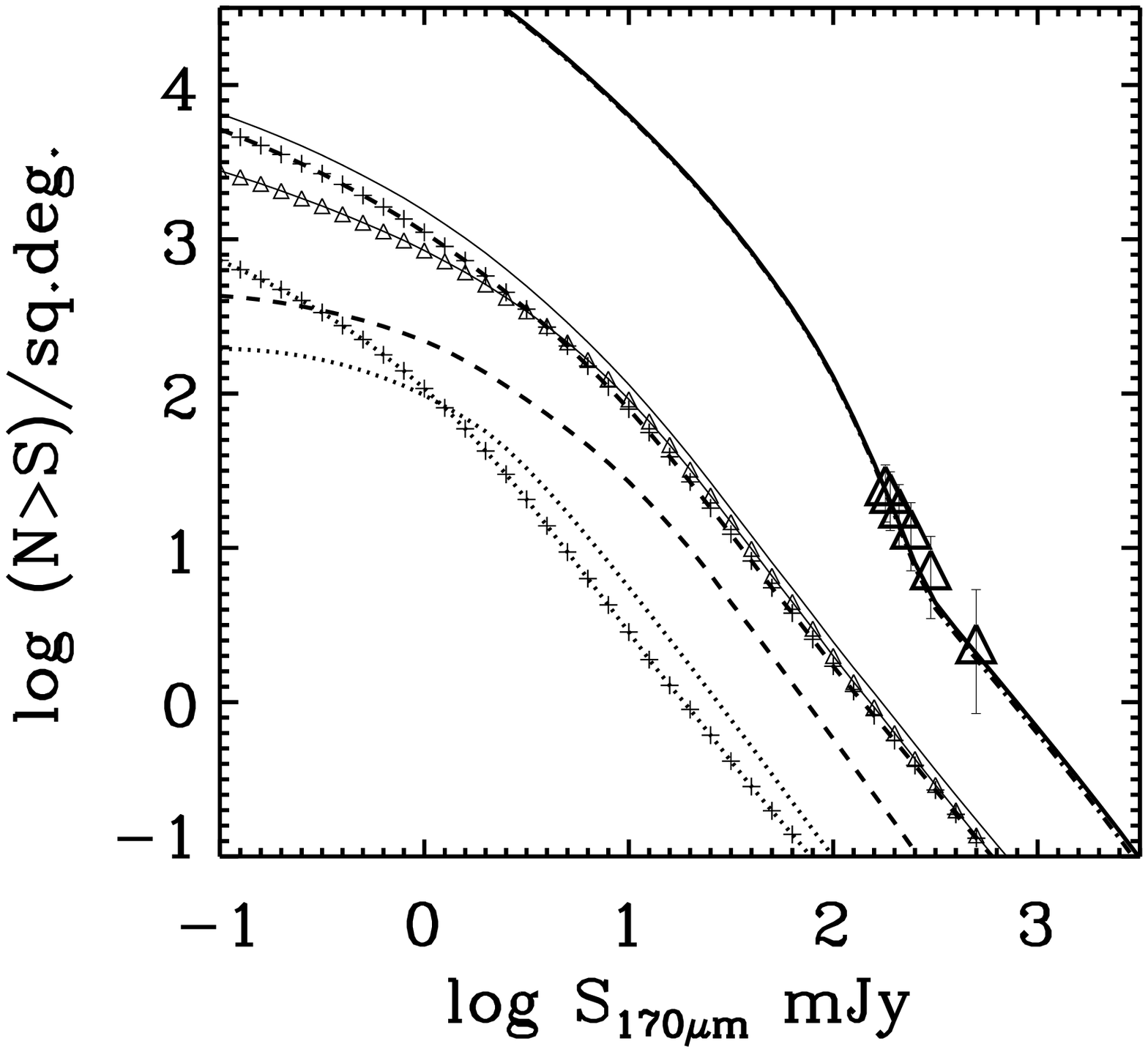}
  \caption{Number counts for AGN+hosts and galaxies in several NIR to FIR bands where data are available.
Meaning of lines is: thin continuous for AGNs + their host
galaxies; dotted and dashed for type 1 and 2 QSOs+ hosts
respectively; the same line styles with plus signs superimposed
are for type 1 and 2 Seyferts+ hosts; the connected triangles show
the contribution by Compton thick AGNs (N$_H>10^{24}$ cm$^{-2}$);
dot-dashed line for galaxies. Data are from: Moustakas et al.
(1997), Kochanek et al. (2001),
  Saracco et al. (2001), Totani et al. (2001), and Cimatti et al. (2002) (K
  band); Taniguchi et al. (1997), Altieri et al. (1999), Serjeant et al.
(2000), Oliver et al. (1997) (closed region with continuous line),
Oliver et al. (2002) (closed region with dashed line), and Sato et
al. (2003) ($6.7\mu$m); Lonsdale et al. (1990), Pearson \&
Rowan-Robinson (1996), Bertin et al. (1997), Mazzei et al. (2001)
($60 \mu$m); Efstathiou et al. (2000), Rodighiero et al. (2003)
($90 \mu$m); Dole et al. (2001) ($170\mu$m).}
              \label{count_ext_dat}
    \end{figure*}

   \begin{figure*}
   \centering
   \includegraphics[angle=0,width=8.5truecm]{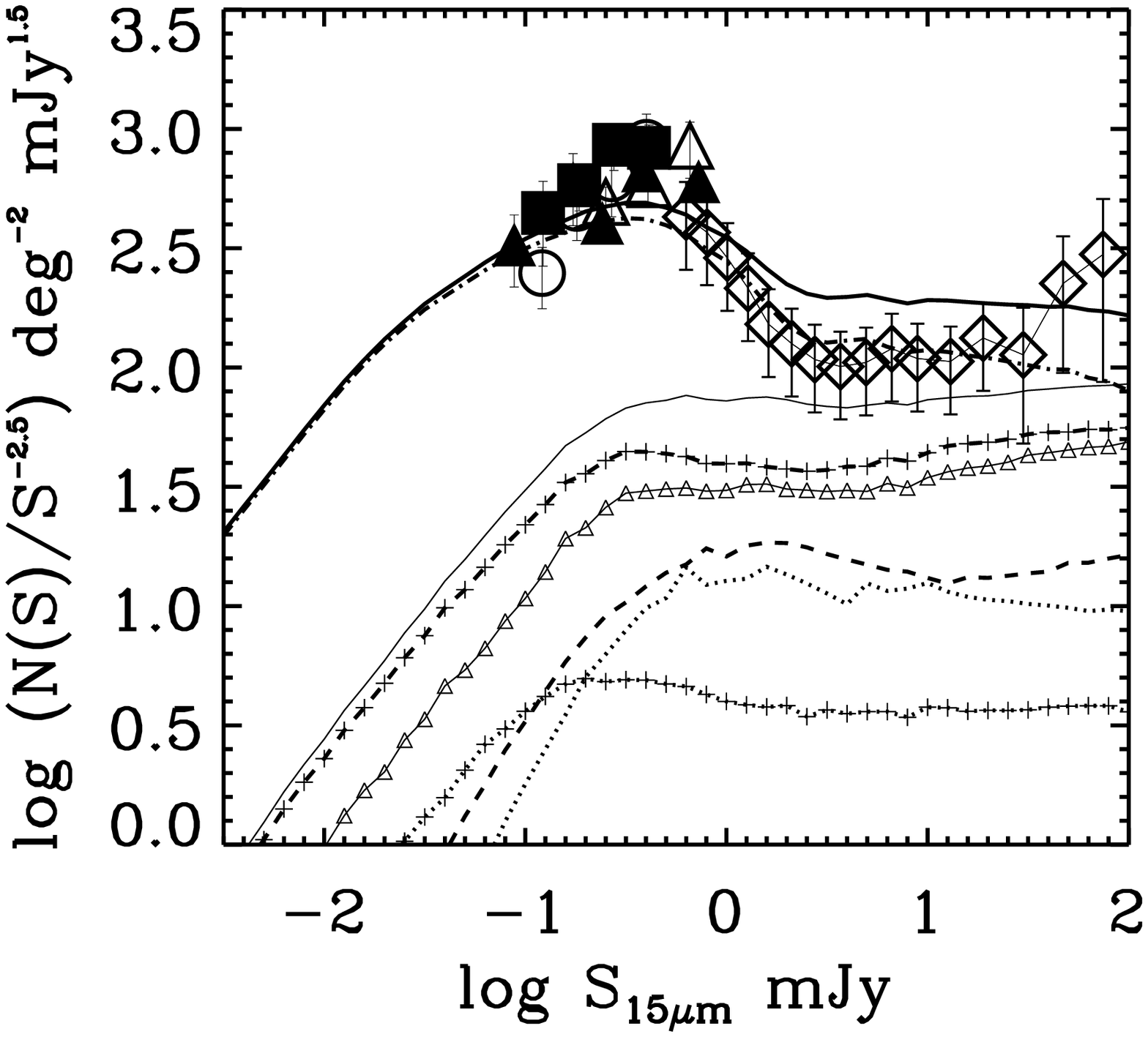}
   \includegraphics[angle=0,width=8.5truecm]{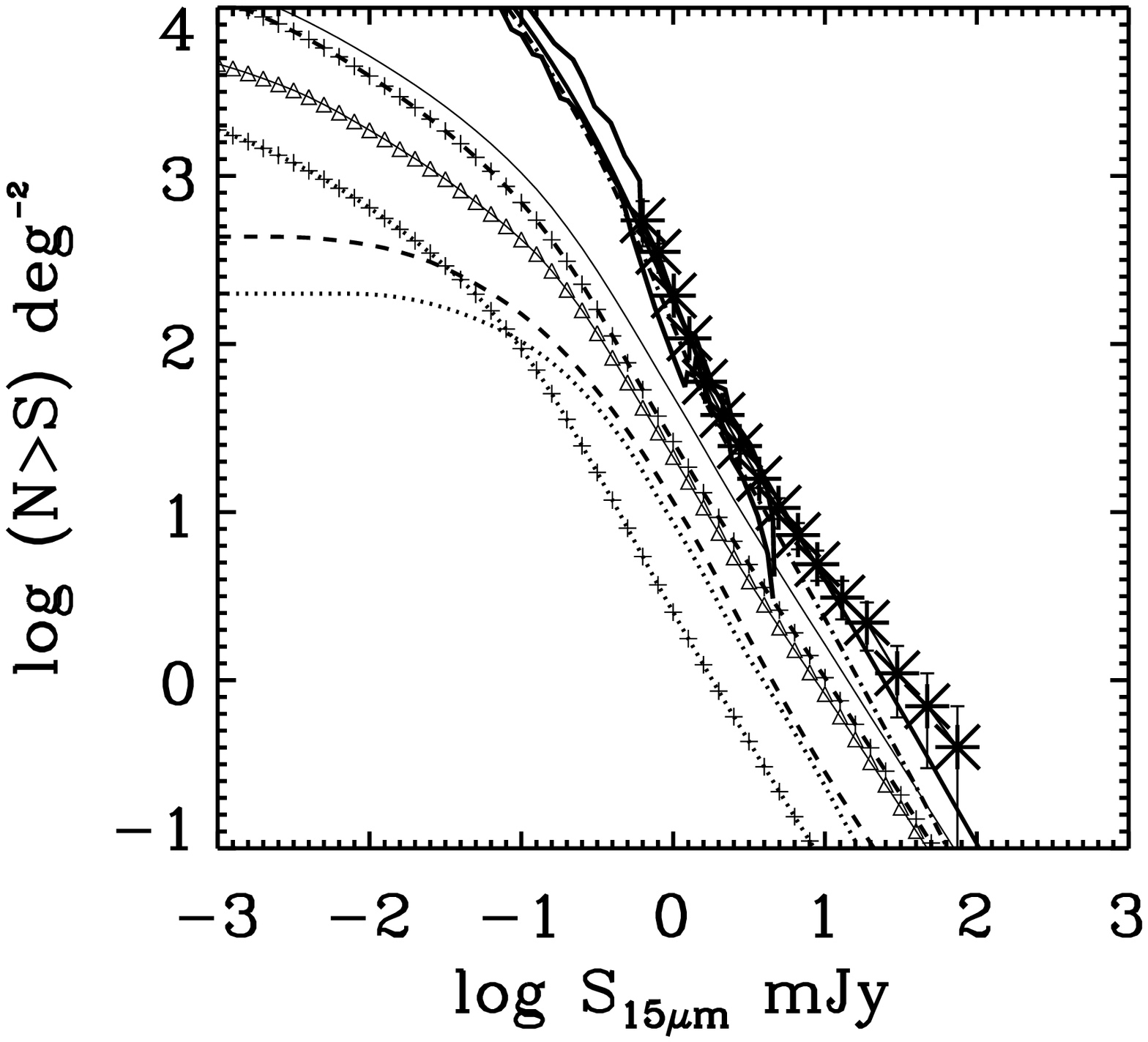}
   \includegraphics[angle=0,width=8.5truecm]{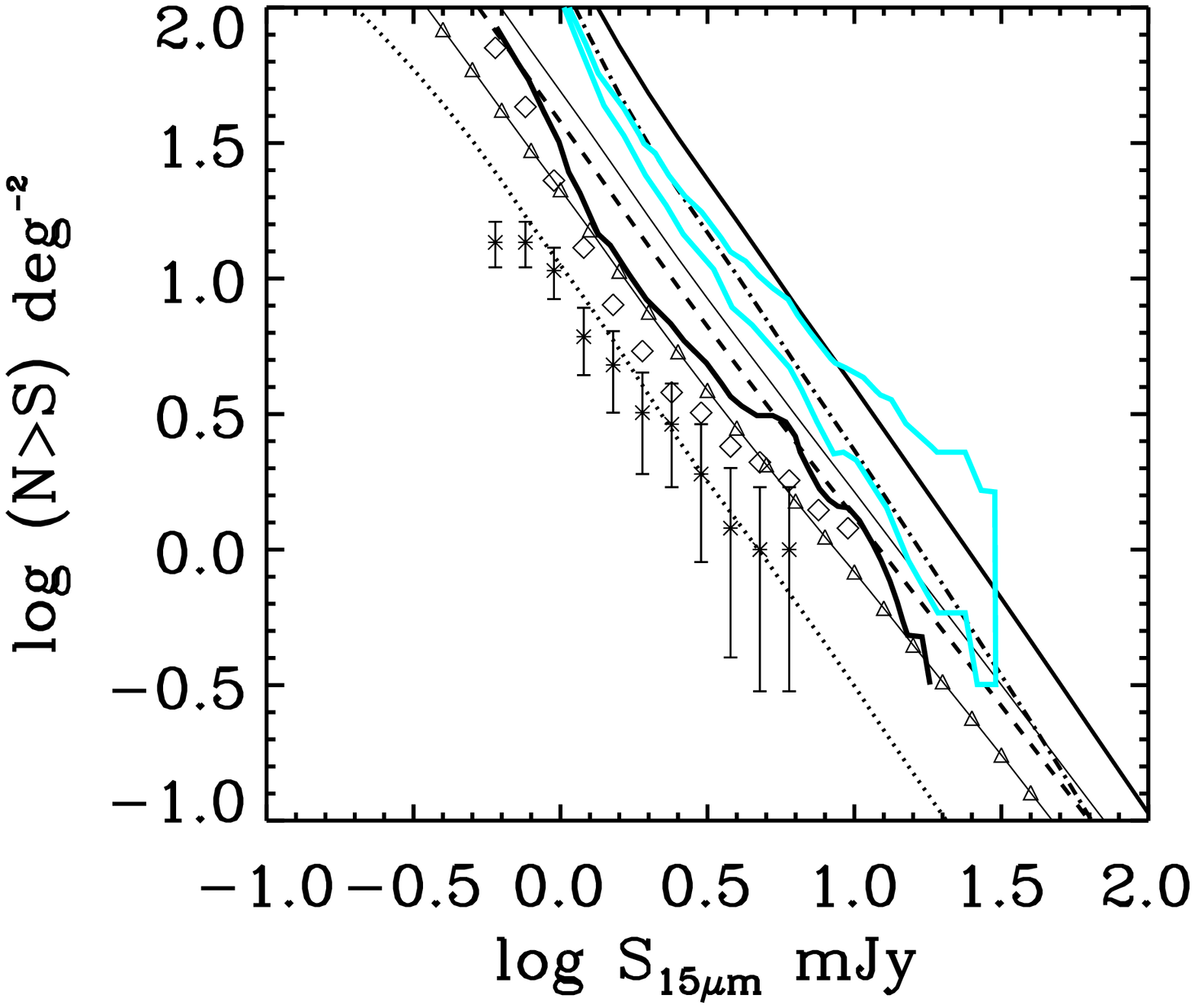}
  \caption{$15 \mu$m number counts for AGN+hosts and galaxies.
  Meaning of lines as in Fig.~\ref{count_ext_dat}. In the upper 2 plots data are from Elbaz et al. (1999)
  and Gruppioni et al. (2002). In the lowest plot, the integral $15 \mu$m counts are compared
with data by La Franca et al. (2004) and La Franca et al. (in
prep.), that have evaluated the source counts due to type 1 and 2
AGNs by means of optical and X-ray identifications of the $15
\mu$m sources. In this plot, data for type 1, type 2 and total
AGNs, and the total counts are shown respectively by asterisks,
diamonds, the thick line, and the the closed region. The
corresponding quantities for our model are shown by the dotted,
dashed, the thin and the thick continuous line. The connected
triangles show the contribution by Compton thick sources.}
              \label{count_ext_15}
    \end{figure*}

When the contribution by the host galaxy is included, the fraction
of the IR background ascribed to AGN+hosts is much higher. This is
shown in Fig. \ref{bg_ext}.

The overall contribution is larger than 5\% at all infrared
wavelengths $\rm \lambda <40 \mu m$, and reaches about
$\gtrsim$10\% for $\lambda \leq 10 \mu$m. The contribution at
15$\mu$m is nearly two times higher with respect to the case of
AGN alone (i.e. AGNs and their host galaxies contribute nearly
equally at this wavelength). The AGN+host contribution decreases
at longer wavelengths, but is still a few percent even in the
sub-mm. The dominant contribution is due to the host galaxies of
Compton thick AGNs.

The comparison of our model with the AGN contribution at 15$\mu$m
observationally inferred by various studies is reasonably good.
The sum of the contribution by type 1 Sys and QSOs at 15$\mu$m
expected by our model matches perfectly the value obtained by
Matute et al. (2002) by the direct identification of the ISO
sources (an identification survey which should be free by
selection effects given that type 1 AGNs are unobscured and
therefore easier to identify). Our estimated total AGN
contribution to the 15$\mu$m background is also close to that
inferred by Fadda et al. (2002), though slightly lower. The slight
difference with Fadda et al. (2002) may be due to a shortage of
Seyfert galaxies at high redshift in the adopted luminosity
function, indeed Ueda et al. (2003) may be slightly incomplete in
this regard since the constraints on the fraction of Seyfert
galaxies at high redshift are looser. We will discuss again this
issue below.

By cross-correlating the Chandra/XMM sources with the submm
sources, various authors have estimated that AGNs contribute for
about 7-15\% to the submm background, depending on the X-ray flux
limit reached (Severgnini et al. 2000; Barger et al. 2001). Our
model predicts a significantly lower contribution ($\sim$3\%) at
850$\mu$m by the host galaxies. Again, this inconsistency is
probably due to a shortage of Seyfert galaxies at high redshift in
the luminosity function by Ueda et al. (2003). This will be
discussed more in detail below when presenting the source number
counts.

   \begin{figure*}
   \centering
   \includegraphics[angle=0,width=8.5truecm]{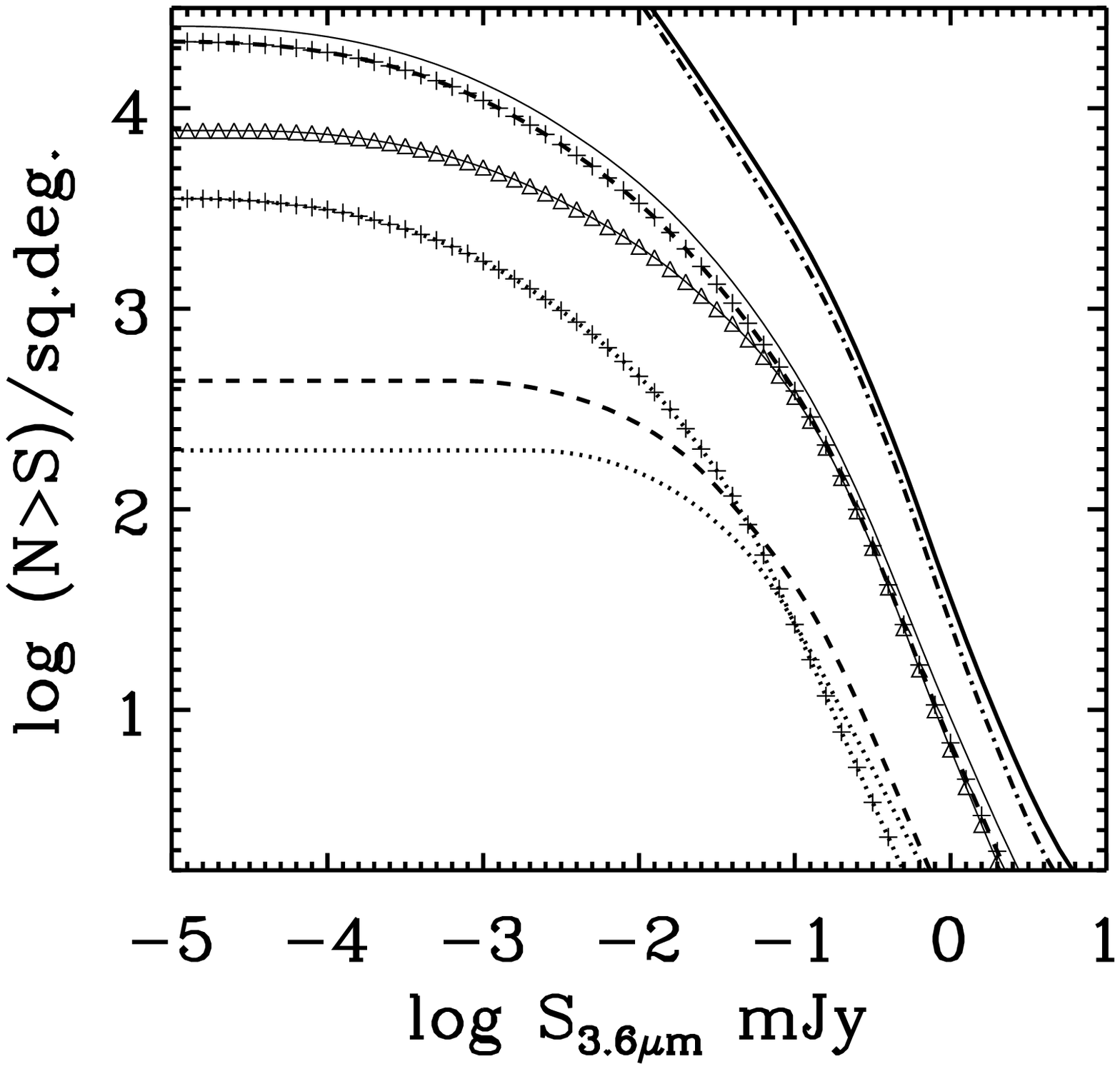}
   \includegraphics[angle=0,width=8.5truecm]{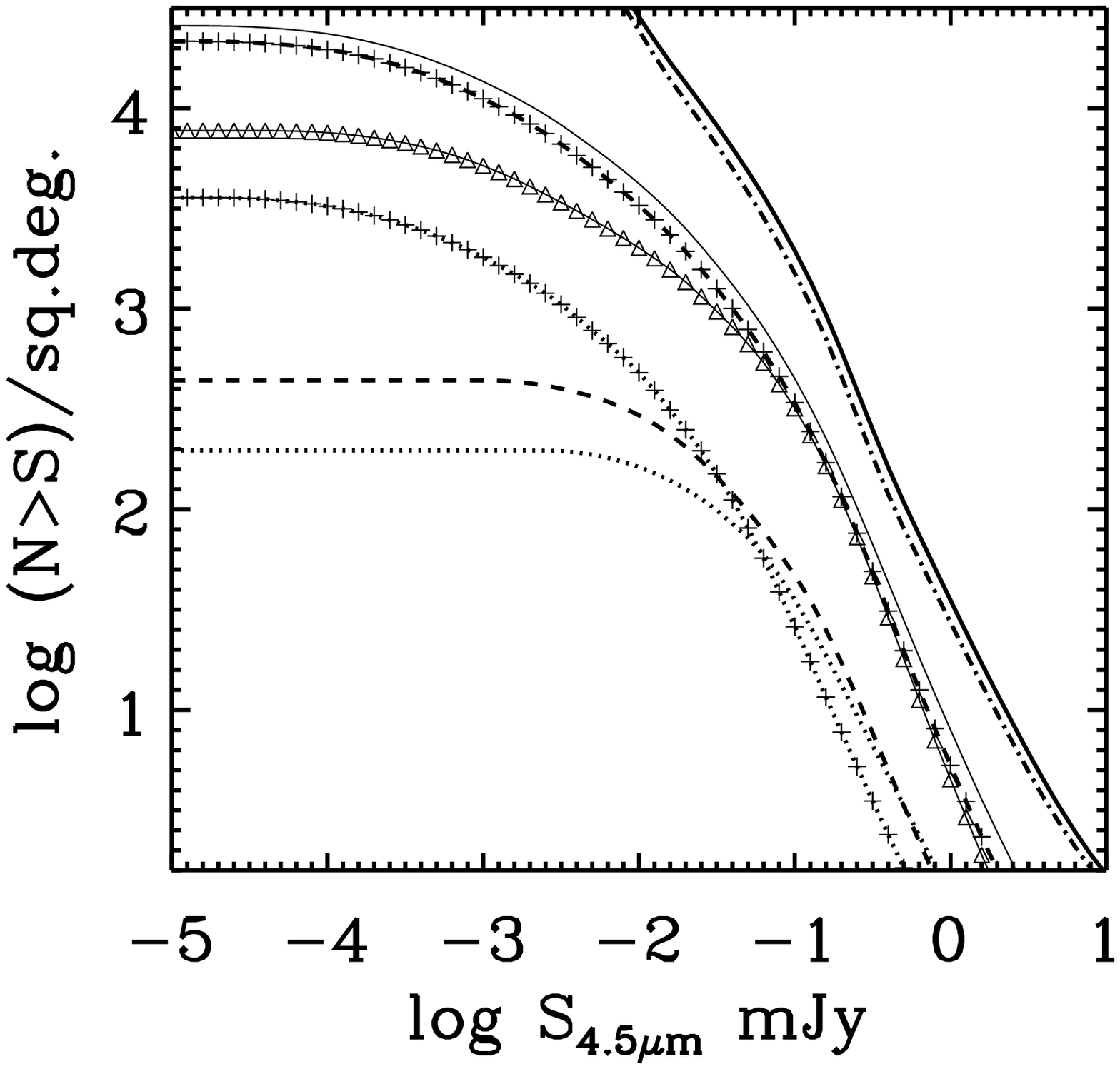}
   \includegraphics[angle=0,width=8.5truecm]{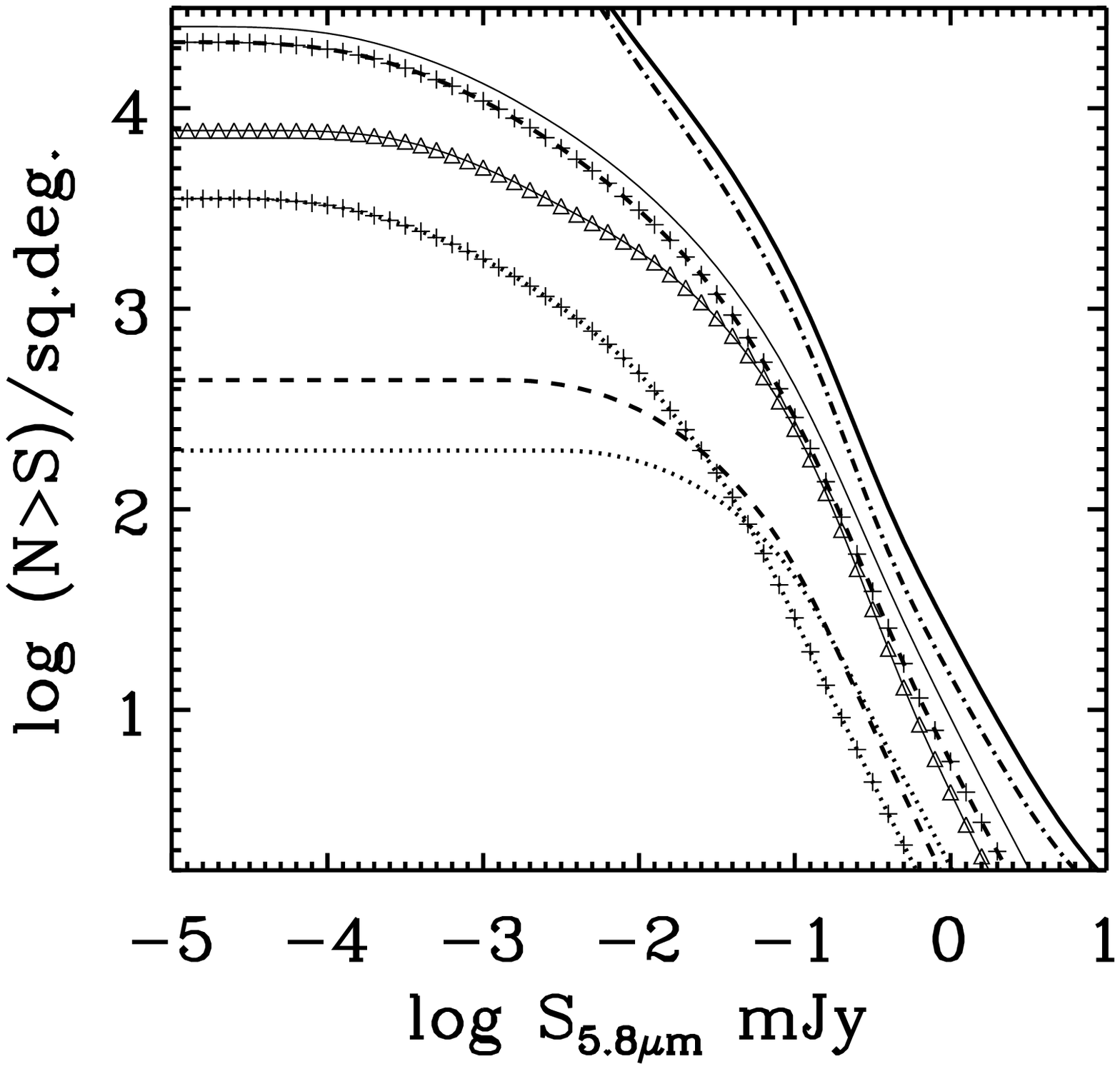}
   \includegraphics[angle=0,width=8.5truecm]{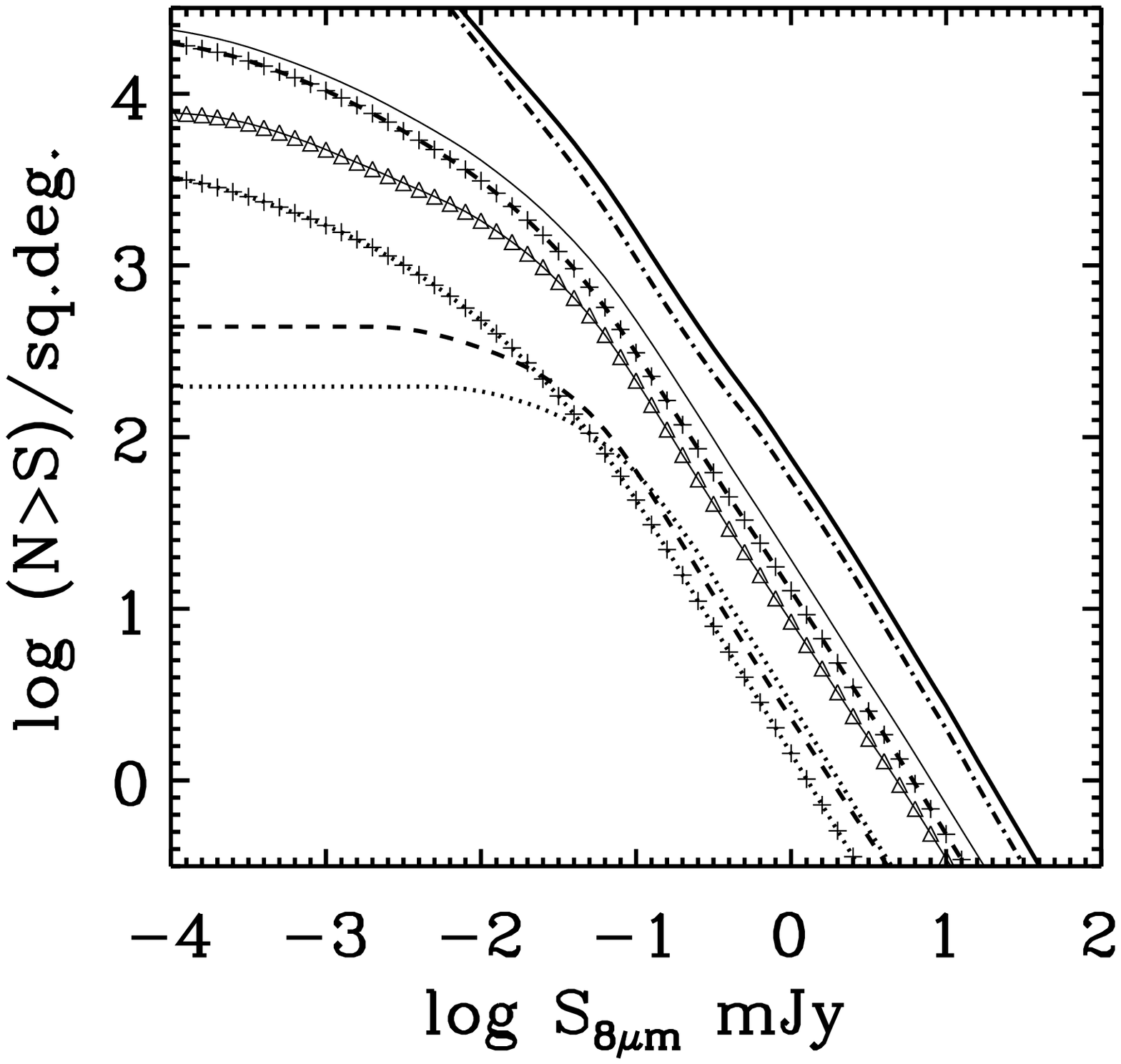}
  \caption{Number counts for AGN+ hosts and galaxies in the SPITZER/IRAC bands.
  Meaning of lines is: thin continuous for AGNs + their host galaxies; dotted and dashed for the
  hosts of type 1 and 2 QSOs respectively; the same line styles with
   plus signs superimposed are for type 1 and 2 Seyferts; the connected triangles show the contribution by Compton
   thick AGNs (N$_H>10^{24}$ cm$^{-2}$); dot-dashed line for galaxies.}
              \label{count_ext_irac}
    \end{figure*}

   \begin{figure}
   \centering
   \includegraphics[angle=0,width=8.5truecm]{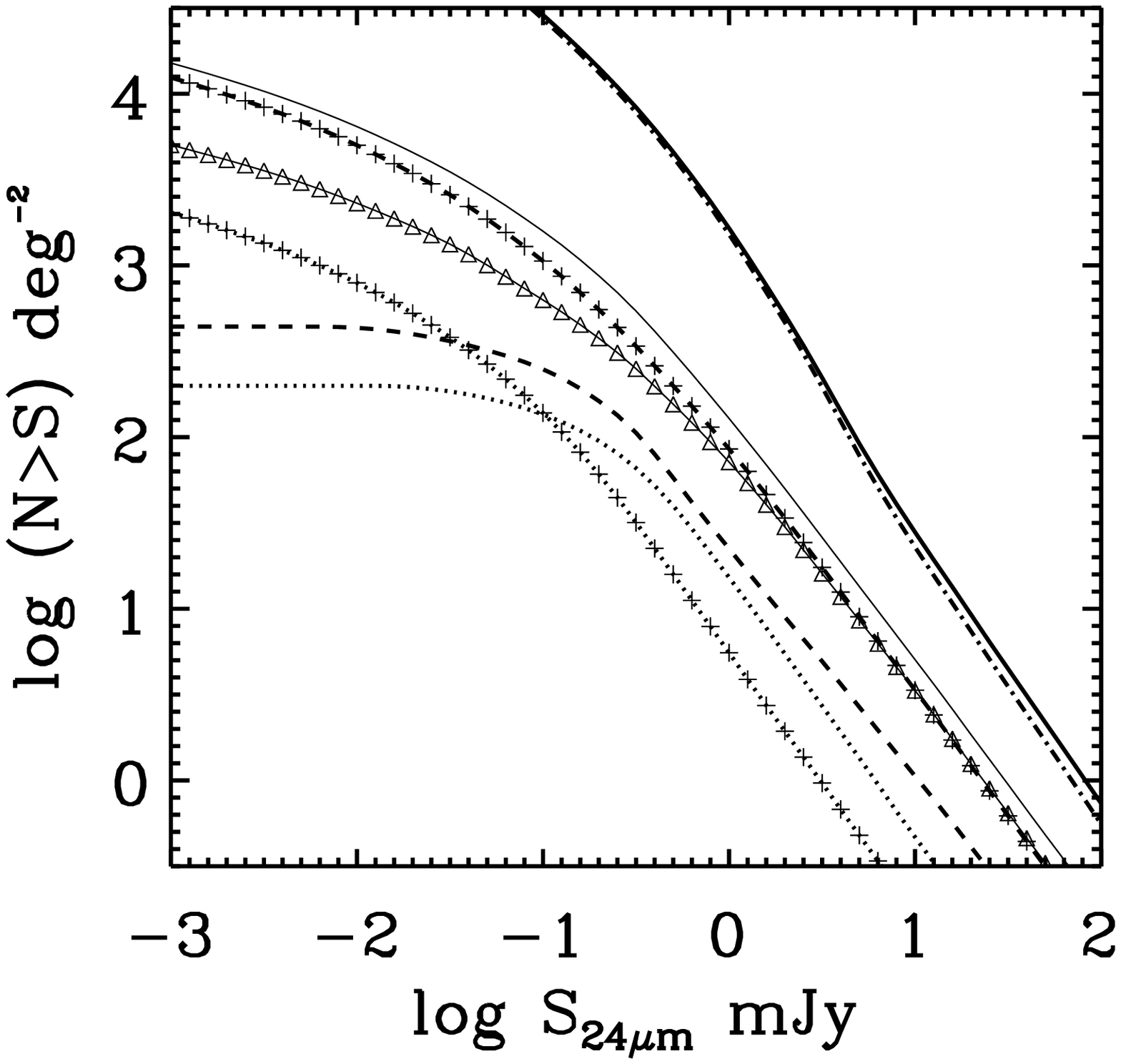}
   \includegraphics[angle=0,width=8.5truecm]{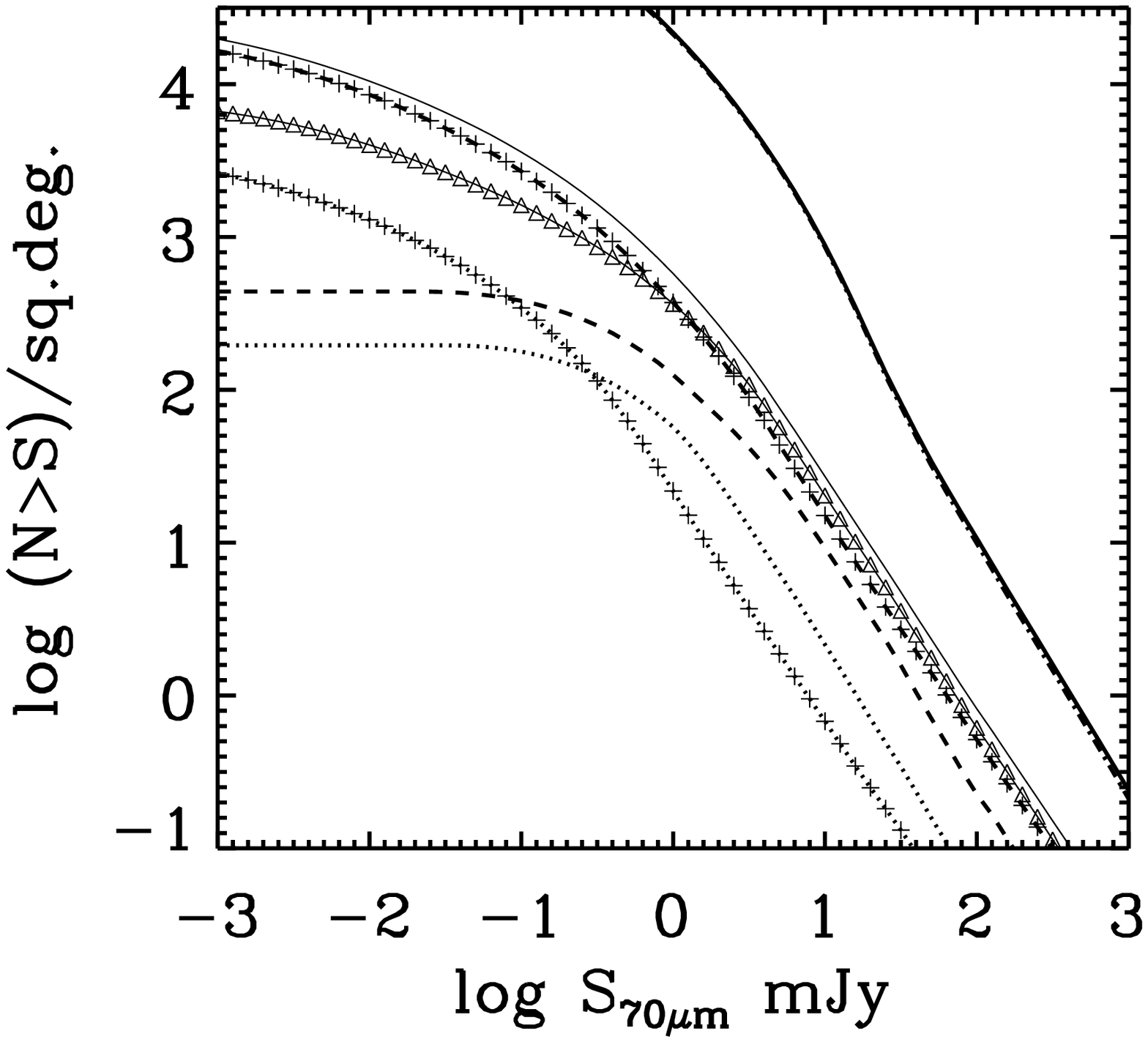}
   \includegraphics[angle=0,width=8.5truecm]{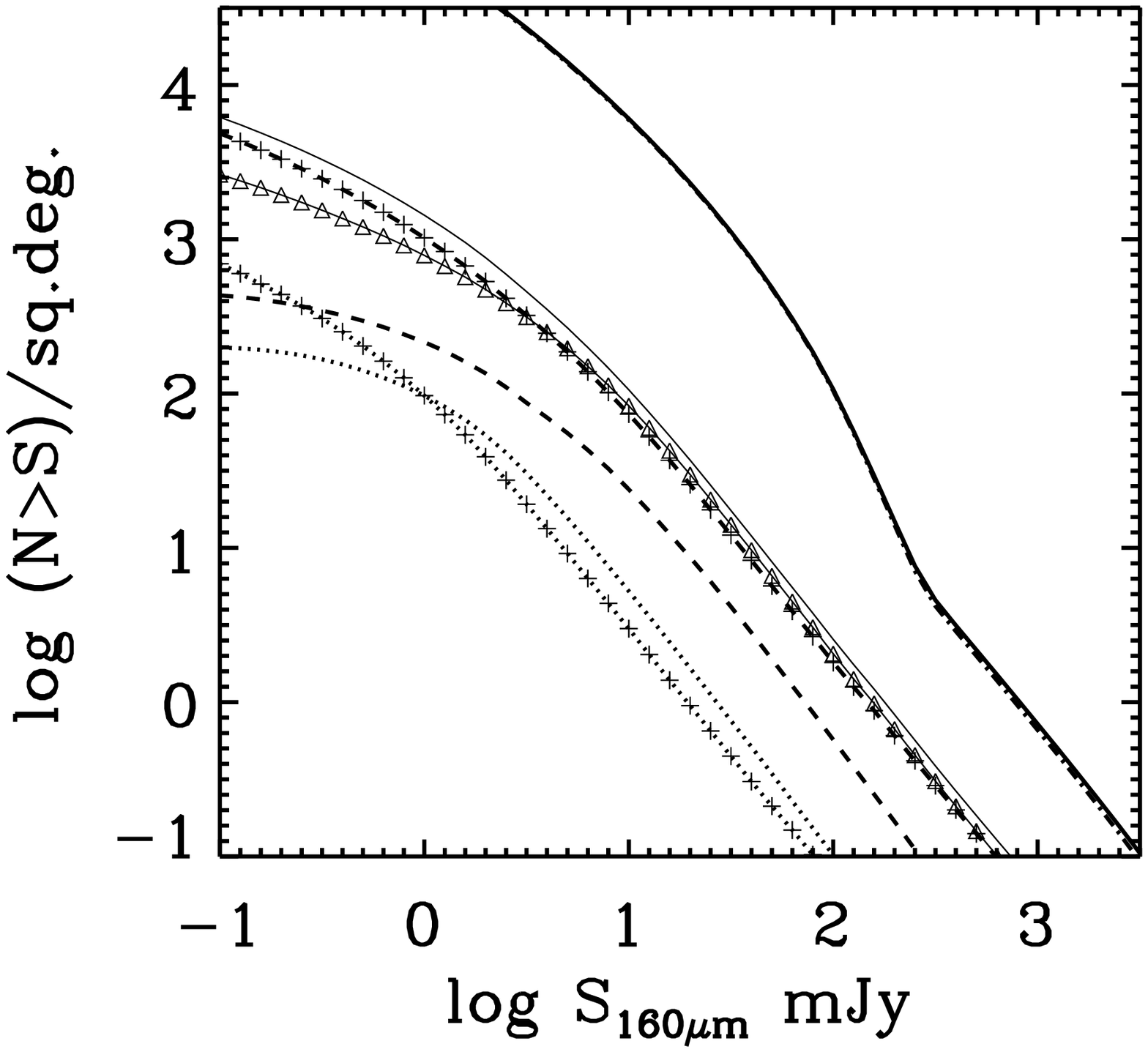}
  \caption{Number counts for AGN+hosts and galaxies in the SPITZER/MIPS bands.
 Meaning of lines is: thin continuous for AGNs + their host
galaxies; dotted and dashed for the
  hosts of type 1 and 2 QSOs respectively; the same line styles with
   plus signs superimposed are for type 1 and 2 Seyferts; the connected triangles show the contribution by Compton
   thick AGNs (N$_H>10^{24}$ cm$^{-2}$); dot-dashed line for galaxies.}
              \label{count_ext_mips}
    \end{figure}

Figs. \ref{count_ext_dat} to \ref{count_ext_mips} show the source
number counts at various infrared wavelengths.

Note that for galaxies we have maintained the same prescriptions
adopted in the previous Section, while in principle we should
decrease their contribution since part of the star formation is
already accounted for through the host galaxies of AGNs. But,
apart a slight overestimate of the bright $15 \mu$m differential
counts (Fig. \ref{count_ext_15}), the models are still within the
available observational constraints, therefore we keep the model
for galaxies without modifications. This can introduce a slight
underestimate of the relative importance of the hosts of AGN at
bright fluxes.

The contribution by AGN+hosts is considerable at all wavelengths
shortward of $\sim$100$\mu$m, exceeding 10\% at intermediate to
bright fluxes ($\rm S\ga 0.1-1$mJy).

In the well studied band at 15$\mu$m  AGN+hosts dominate the
counts at fluxes higher than a few mJy (the AGN+hosts contribution
is $\sim 10$\%, 30\%, 40\% and 60\% respectively at flux levels of
$0.1$, 1, 5 and 100 mJy, Fig. \ref{count_ext_15}). Qualitatively
our finding is in agreement with the recent study by La Franca et
al. (2004) whose optical identification of the ISO 15$\mu$m
sources (combined with their preliminary results on X-ray
identifications, La Franca et al., in prep.) indicate a rapidly
increasing fraction of AGNs at high fluxes. Quantitatively La
Franca et al. (2004) find only half of the AGNs expected by our
model, but the statistical errorbars are actually consistent with
our values at the 1$\sigma$ level. However, if one wishes to
investigate this marginal inconsistency, one possibility, as
suggested by La Franca et al. (2004), could be that optical
spectroscopic identifications have missed a significant fraction
of type 2 AGNs. Our model indicates that most of the AGNs
contributing to the source counts are heavily obscured (Compton
thick) and therefore are more likely to be missed by optical
identification. In particular, Maiolino et al. (2003) have clearly
shown that several Compton thick AGN have been missed by optical
spectroscopy even in nearby galaxies. La Franca et al. (in prep.)
attempt to account for such optically elusive AGNs, by exploiting
the current results on this regard from X-ray surveys. Nonetheless
as pointed out both by Ueda et al. (2003) and by Maiolino et al.
(2003) most of such Compton thick AGN are probably missed even by
the latest hard X-ray surveys. Another possibility, as suggested
by the comparison with La Franca et al. (2004), is that the
extrapolation of the luminosity functions by Ueda et al. (2003)
tend to overestimate the density of AGNs at low redshift.

A much more critical issue is the contribution of AGN to the
number counts at 850$\mu$m. Alexander et al. (2003) have found
that about half of the submm sources brighter than 5~mJy host a
Seyfert nucleus. In a more recent work, Alexander et al. (2004)
identify most of these sources as galaxies at z$\sim$2 (i.e. at a
redshift comparable with most of the submm sources at this
limiting flux). Our model predicts that only a few percent of the
submm sources host an AGN at fluxes $\rm S_{850\mu m}> 5$~mJy.
Again, the main problem seems to be that the luminosity functions
of Ueda et al. (2003) underestimate the number of Seyfert galaxies
at high redshift.

Indeed, at z$>$2 the luminosity function at Seyfert luminosities
is not tightly constrained by the observational data available to
Ueda et al. (2003) and therefore it is mostly extrapolated from
lower redshifts. This issue may also affect our estimated
contribution of AGN hosts to the far-IR bands of Spitzer. A more
detailed investigation of this issue and of the possible
implications for the Seyferts evolution at high redshift will be
discussed in an upcoming paper (Silva et al. in prep.). An
additional problem might be our modelling of the hosts SED in the
submm. Indeed, due to the lack of submm data for most of the
sources in our sample, the long wavelength tail of the SEDs in
Fig. \ref{host_sed} are mostly obtained by extrapolating the
far-IR data with theoretical models which fit properly the SEDs of
local starburst and quiescent galaxies. However, such
extrapolation may be wrong by some factor.

In the Spitzer bands with $\lambda \leq 24 \mu$m we expect that
important fractions of the total detected sources will be provided
by galaxies hosting an AGN. This is particularly true at bright
flux limits. At the sensitivity limits reported by Lonsdale et al.
(2003) for the SWIRE suvey ($7.3$, $9.7$, $27.5$, $32.5$ $\mu$Jy
respectively for the IRAC $3.6$, $4.5$, $5.8$, $8$ $\mu$m bands,
and $0.45$ mJy for the $24 \mu$m band of MIPS), we get that $\sim
7$ to $25$\% of the total sources host AGNs, with the maximum
value at $5.8$ and $8 \mu$m.

At the longer MIPS bands, $70$ and $160 \mu$m, at flux limits in
the ranges  $\sim 5$--10~mJy, and $\sim 70$--100~mJy respectively,
we expect that only a few percent of the detected sources will
host an AGN. However, at these long wavelengths our estimates may
be affected by the same problem as for the submm sources as a
consequence of the shortage of high-z Seyferts in the Ueda et al.
(2003) model.

\section{Comparison with previous studies}

In the previous Section we have compared the predictions of our
model with observational constraints at specific wavelengths.
However, the contribution of AGNs to the IR background and IR
number counts has been modelled and investigated by other authors
in the past. In this Section we compare our results with some of
these previous studies.

In Granato et al. (1997), predictions for the contribution of AGNs
to the IR background and MIR source counts were presented,
starting from the X-ray background as a constraint. Theoretical
nuclear SEDs as a function of $N_H$ were used for AGNs, and the
contribution by the host galaxy was considered in a very
approximate manner. Their results are within a factor of 2 as
compared to ours for the nuclear emission by AGNs in the $\sim
6-90 \mu$m range. The reason is that they assumed that the MIR
SEDs are ascribed to the AGN only.

Starting from their observed contribution to the 15$\mu$m
background from type 1 AGN, Matute et al. (2002) infer the total
contribution from all AGN to the 15$\mu$m background. Under
``standard'' assumptions based on the unified theories they infer
a total contribution of $\sim$10\%--15\%, which is similar to that
inferred by us.

Risaliti, Elvis, \& Gilli (2002) predict that the maximum
contribution from AGNs to the infrared background occurs at
60$\mu$m and at the level of 20\%--40\%. This is significantly
different from what predicted by our model (which predicts a
maximum contribution by AGN+hosts at about 10$\mu$m and not
exceeding 20\%. However, there are two main problems in Risaliti
et al. (2002). First, they assume the bolometric correction (the
proportionality between $\rm L_X$ and $\rm L_{bol}$) typical of
Quasars, while it is now known that a large fraction of the X-ray
background is produced at Seyfert-like luminosities (for which the
bolometric correction is lower by a factor of about three).
Secondly, they assume that most of the quasars are absorbed
(80\%-90\%), while recently it has become clear that at quasar
luminosities the fraction of obscured AGN is much lower than for
Seyfert, and more specifically the fraction of obscured AGN at
$\rm L_X>10^{44}~erg~s^{-1}$ is only 30\%-40\%. These two factors
are mostly responsible for the excess of IR background ascribed to
AGNs in Risaliti et al. (2002).

More recently Andreani, Spinoglio, \& Malkan (2003) have derived
the contribution of various classes of active galaxies to the
source number counts in different infrared bands. They consider
separately the contribution by Sy1s, Sy2s and QSO1s (but not
QSO2s) and use the SEDs inferred by  Spinoglio et al. (2002) which
include the contribution from the host galaxies. Therefore, their
predictions should be comparable with our results in Section
\ref{sec:agnhost}. They adopt pure luminosity evolutions from
Malkan \& Stecker (2001) and Boyle et al. (2000). It is now clear
that AGNs follow a Luminosity Dependent Density Evolution (LDDE),
however pure luminosity evolution formulas provide a reasonable
approximation for the behavior of the bulk of AGNs around the knee
of the luminosity function and above. The predictions on the
source number counts obtained by Andreani et al. (2003) in the
various IR bands are close to those obtained by us, at least for
what concerns the {\it total} AGN contribution. Yet, our model
predicts a larger contribution by obscured (type 2) AGN than
predicted by Andreani et al. (2003). On this regard, one should
keep in mind that the latter did not include the contribution by
type 2 quasars.

\section{Conclusions}

We have estimated the contribution to the infrared background and
infrared number counts from AGNs by using the most recent AGNs
luminosity functions and evolution from the hard X-ray surveys
(Ueda et al. 2003) and by connecting their X-ray and IR emission
by means of new detailed AGN SEDs derived by us. This approach
allow us to derive the contribution of AGN to the infrared
background by using mostly observed and measured quantities, with
little additional assumptions.

The main results are the following:
\begin{enumerate}
\item The AGN nuclear IR radiation contributes little ($<$5\%) to
the IR background at most wavelengths.
\item AGNs along with their
host galaxies contribute significantly ($\sim$~5--20\%) to the IR
background at $\lambda < 60 \mu m$.
\item AGNs and their hosts
probably dominate the mid-IR number counts at bright fluxes
($>$1~mJy).
\item The predicted contribution of AGNs and of their
host galaxies
 to the mid-IR background and number counts are in good agreement
 with the observational constraints available so far.
\item However, we find a strong discrepancy between the expected
contribution of Seyfert host galaxies at high redshift to
the submm background and number counts at bright fluxes. We ascribe
this discrepancy mostly to the shortage of Seyferts at high redshift
(z$\sim$2) in the X-ray luminosity functions
which, at these high redshifts and low luminosities,
are still poorly constrained by the current hard X-ray surveys.
This issue may also yield to underestimate the AGN hosts contribution
to the far-IR bands of Spitzer. A more detailed investigation
of this issue and of the possible implications for the Seyferts evolution
at high redshift will be discussed in an upcoming paper (Silva et al. in prep.).
\end{enumerate}

\section*{Acknowledgments}

This work was partially supported by the Italian Ministry of
Research (MIUR), by the National Institute for Astrophysics (INAF)
and by the Italian Space Agency (ASI). LS and GLG acknowledge kind
hospitality by INAOE where part of this work was performed, and
SAGG for partial financial support. We thank Fabio La Franca for
useful discussions and for providing the 15 $\mu$m data in advance
of publication. We thank the referee for careful reading and for
many useful comments that helped us to improve the presentation.

\appendix

\section{The samples used to derive
spectral energy distributions}

In Tables A1, A2 and A3 we list the samples of Sy1s, Sy2s and
QSOs, respectively, used to derive the IR SEDs discussed in
Sections 3 and 4. The tables report the {\it intrinsic} hard X-ray
luminosities (i.e. corrected for absorption), the nuclear IR
luminosities inferred by our nuclear SEDs, and the total IR
luminosities (i.e. AGN+host). For Sy1s and QSOs we also report the
luminosity of the UV-optical bump inferred by adapting the
combined average SED of type 1 AGN obtained by Vanden Berk et al.
(2001) and by Telfer et al. (2002) (the latter for the extreme UV
part of the spectrum) to the optical nuclear photometric points
(either from HST or from high resolution groundbased optical
observations).

The sample and IR SED of Compton thick Sy2s with $\rm
10^{24}<N_H<10^{25}cm^{-2}$ deserves some more discussion. Near
infrared observations of this class of objects shows more rarely
indication of non stellar emission, strongly suggesting that heavy
dust obscuration in these objects often prevents the detection of
the nuclear near-IR source. Of the 5 Sy2s in this $\rm N_H$ range
only two (namely NGC3281 and Circinus) show clear evidence for
near-IR non stellar emission. For these we could fit the Granato
\& Danese (1994) models to the data and derive their nuclear IR
SED. However, most likely these two objects represent the less
absorbed cases of this class, therefore providing a biased view
their SED. Therefore we tried to account also for the SED of the
more obscured AGNs. In the other three sources with $\rm
10^{24}<N_H<10^{25}cm^{-2} $ (namely NGC4945, NGC3079, NGC5194)
there is probably a detection of the obscured AGN only at 10$\mu$m
(e.g. Krabbe, Boker, \& Maiolino 2001). We used the IR SED
determined for Circinus and NGC3281, normalized to the X-ray
intrinsic flux of the three sources and then increased the
obscuration of the Granato \& Danese (1994) model (by increasing
the inclination angle of the torus) until the SED matched the
observed 10$\mu$m nuclear photometry, with the additional
constraint that the integrated nuclear IR emission should not
change. This is a rough method to include the SED of the very
obscured AGN with $\rm 10^{24}<N_H<10^{25}cm^{-2}$, but it is the
best guess that can be made with the available data.

All the average SEDs obtained in this paper are available
in electronic form at the following web site:
{\sl http://www.arcetri.astro.it/$\sim$maiolino/agnsed}.

   \begin{table*}
      \caption[]{Sample of Seyfert 1 galaxies}

         %\begin{array}{p{0.5\linewidth}l}
         \begin{tabular}{lccccccc}
            \hline
            \noalign{\smallskip}
        Name      &  z & lg L$_{\rm 2-10keV}^a$ & lg L$_{\rm IR}$(nuc)$^{a,b}$ &
     lg L$_{\rm IR}$(tot)$^{a,c}$ & lg L$_{\rm UV-opt}$(nuc)$^{a,d}$ &
     Refs.(X)$^e$ & Refs.(IR)$^f$\\

           \noalign{\smallskip}
            \hline
            \noalign{\smallskip}

   NGC 3227    &  0.0039 &  41.9  &  42.3 &   42.6 &  43.5 & 1 & 1,2,3 \\
   NGC 3516    &  0.0088 &  43.1  &  42.7 &   43.3 &  43.8 & 1 & 4 \\
   NGC 3783    &  0.0097 &  43.0  &  43.2 &   43.4 &  43.9  & 1 & 5,6,7 \\
   NGC 4051    &  0.0023 &  41.3  &  41.9 &   42.1 &  42.0 & 1 & 4 \\
   NGC 4151    &  0.0033 &  42.7  &  43.0 &   42.4 &  43.2 & 2 & 1,8 \\
   NGC 4253    &  0.0129 &  42.2  &  43.2 &   43.6 &  44.0 & 3 & 4 \\
   NGC 4593    &   0.009 &  42.8  &  42.9 &   43.3 &  43.1 & 4 & 9,10,5,11,12,8 \\
   NGC 5548    &  0.0172 &  43.4  &  43.5 &   43.9 &  44.4 & 4 & 1,13,9 \\
   NGC 7213    &  0.0039 &  42.1  &  42.2 &   42.6 &  43.1 & 5 & 10,14,15 \\
   NGC 7469    &  0.0163 &  43.3  &  44.2 &   43.8 &  45.0  & 4 & 1,2,8 \\
   IC 4329     &  0.016  &  43.8  &  44.0 &   43.8 &  45.2 & 1 & 1,8 \\
   Mrk 1095    &  0.0323 &  44.0  &  43.8 &   44.4 &  --   & 3 & 5,6,8 \\
   Mrk 1513    &  0.0629 &  43.6  &  44.4 &   45.0 &  45.5 & 6 & 4 \\
   Mrk 335     &  0.0258 &  43.1  &  43.6 &   44.2 &  44.7 & 1 & 4 \\
   Mrk 590     &  0.0263 &  43.6  &  43.5 &   44.2 &  44.1 & 4 & 4 \\
   Mrk 841     &  0.0129 &  42.6  &  43.2 &   43.6 &  43.6 & 7 & 4 \\
   MCG-6-30-15 &  0.0077 &  42.8  &  42.9 &   43.2 &  43.4 & 1 & 5,8,11 \\
            \noalign{\smallskip}
            \hline
         \end{tabular}
\begin{flushleft}
$^a$ All luminosities are in units of erg~s$^{-1}$ \\
$^b$ Nuclear infrared luminosity associated
   with dust emission heated by the AGN calculated in the range 8--1000$\mu$m (for
   comparison in $\rm L_{IR}$ defined by Sanders \& Mirabel (1996).\\
$^c$ Total (AGN+host)
infrared luminosity in the range 8--1000$\mu$m estimated by using the IRAS data and
following the prescription in Sanders \& Mirabel (1996).\\
$^d$ Luminosity of the UV-optical blue bump of the AGN (see text).\\
$^e$ References for the X-ray data: 1--Reynolds (1997), 2--Malizia
et al. (1997), 3--Turner et al. (1999), 4--Perola et al. (2002),
5--Imanishi \& Ueno (1999), 6--Lawson \& Turner (1997),
7--Ceballos \& Barcons (1996).
\\
$^f$ References for the nuclear IR data: 1--Alonso-Herrero et al.
(2001), 2--Krabbe, Boker, \& Maiolino (2001), 3--Wright et al.
(1988), 4--Alonso-Herrero et al. (2003), 5--Oliva et al. (1999),
6--McAlary et al. (1983), 7--Frogel, Elias, \& Phillips (1982),
8--Ward et al. (1987), 9--Quillen et al. (2001) 10--Kotilainen et
al. (1992) 11--Glass \& Moorwood (1985) 12--Maiolino et al. (1995)
13--Rieke (1978) 14--Ward et al. (1982) 15--Glass, Moorwood, \&
Eichendorf (1982)
\\
 \label{tab_sy1}
\end{flushleft}
   \end{table*}

   \begin{table*}
      \caption[]{Sample of Seyfert 2 galaxies}

         %\begin{array}{p{0.5\linewidth}l}
         \begin{tabular}{lccccccc}
            \hline
            \noalign{\smallskip}
        Name      &  z & $\rm N_H$$^a$ &lg L$_{\rm 2-10keV}^{b,c}$ &
    lg L$_{\rm IR}$(nuc)$^{b,d}$ &
     lg L$_{\rm IR}$(tot)$^{b,e}$ &
     Refs.(X)$^f$ & Refs.(IR)$^g$\\

           \noalign{\smallskip}
            \hline
            \noalign{\smallskip}
 %   NGC 1068    & 0.00379  &$>$1000. & --   & 44.2     &  44.9    & 1 & 1,2,3,4  \\
     NGC 1365    & 0.005457 &   20$^{+4}_{-4}$   & 42.1 & 42.6  &     44.8 & 2 & 5,6,7,8  \\
     NGC 3079    &   0.006  & 1000$^{+540}_{-530}$ & 42.5 & 42.6    &   44.6   & 3 & 15 \\
     NGC 3081    & 0.007955 &   64$^{+20}_{-12}$   & 42.4 & 42.7 &   42.1 & 4 & 10,11,12,13 \\
     NGC 3281    & 0.01067  &  196$^{+19}_{-5}$  & 43.2 & 43.6     &  44.4    & 5 & 14,9,13  \\
     NGC 4388    & 0.008419 &   42$^{+6}_{-10}$ & 42.8 & 42.6  &    44.2 & 2 & 10,11,16,17 \\
     NGC 4945    & 0.00093  &  400$^{+20}_{-12}$  & 41.6 & 41.5     &  43.9    & 6 & 9 \\
     NGC 5194    & 0.00093  &  560$^{+40}_{-16}$  & 40.6 & 40.6     &  43.2    & 7 & 20 \\
     NGC 5252    & 0.00618  &  4.3$^{+0.6}_{-0.6}$ & 41.9 & 41.7   &    41.9  & 2 & 1,21  \\
     NGC 5506    & 0.00618  & 3.4$^{+0.3}_{-0.1}$ & 42.9 & 43.1   &    44.0  & 2 & 1,18  \\
     NGC 5674    & 0.0249   & 7.0$^{+2.8}_{-2.6}$   & 43.2 & 43.1 &  44.4  & 2 & 10,16,19,20\\
     NGC 7172    & 0.00868  & 8.6$^{+0.8}_{-0.3}$  & 42.5 & 42.8   &  44.0  & 2 & 1,21  \\
     NGC 7582    & 0.00528  & 12.4$^{+0.6}_{-0.8}$ & 42.5 & 42.9  &   44.4 & 2 & 10,22,5,6,7\\
     Mrk 348     &  0.015   & 10.6$^{+3.1}_{-2.6}$ & 43.0 & 43.5   &    44.0  & 2 & 1,23  \\
     Circinus    & 0.00093  & 430$^{+40}_{-70}$  & 41.9 & 42.6    &  43.7   & 8 & 10,24,25,9\\
     MGC-5-23-16 &  0.0082  & 1.6$^{+0.2}_{-0.2}$  & 43.1 & 43.3   &  42.2    & 2 & 1,7 \\
            \noalign{\smallskip}
            \hline
         \end{tabular}
\begin{flushleft}
$^a$ Intrinsic absorbing column density in units of $\rm 10^{22}~cm^{-2}$.\\
$^b$ All luminosities are in units of erg~s$^{-1}$ \\
$^c$ The hard X-ray lumionsity is corrected for absorption.\\
$^d$ Nuclear infrared luminosity associated
   with dust emission heated by the AGN calculated in the range 8--1000$\mu$m (for
   comparison in $\rm L_{IR}$ defined by Sanders \& Mirabel (1996).\\
$^e$ Total (AGN+host)
infrared luminosity in the range 8--1000$\mu$m estimated by using the IRAS data and
following the prescription in Sanders \& Mirabel (1996).\\
$^f$ References for the X-ray data: 1--Matt et al. (1997),
2--Bassani et al. (1999), 3--Iyomoto et al. (2001), 4--Maiolino et
al. (1998b), 5--Vignali \& Comastri (2002), 6--Guainazzi et al.
(2000), 7--Fukazawa et al. (2001), 8--Matt et al. (1999).
\\
$^g$ References for the nuclear IR data: 1--Alonso-Herrero et al.
(2001), 2--Tresch-Fienberg et al. (1987), 3--Papadoupoulos \&
Seaquist (1999), 4--Schinnerer et al. (2000), 5--Oliva et al.
(1999), 6--McAlary et al. (1983), 7--Frogel et al. (1982),
8--Deveraux (1989), 9--Krabbe et al. (2001), 10--Quillen et al.
(2001) 11--Alonso-Herrero et al. (1998), 12--Ward et al. (1982)
13--Boisson \& Durrett (1986) 14--Simpson (1998)
15--Alonso-Herrero et al. (2003), 16--McLeod \& Rieke (1994a),
17--Scoville et al. (1983), 18--Ward et al. (1987), 19--Ivanov et
al. (2000), 20--Maiolino et al. (1995), 21--Aitken \& Roche
(1984), 22--Kotilainen et al. (1992) 23--Rieke (1978) 24--Maiolino
et al. (1998a) 25--Siebenmorgen et al. (1997) 26--Glass et al.
(1982)
\\
 \label{tab_sy2}
\end{flushleft}
   \end{table*}

   \begin{table*}
      \caption[]{Sample of quasars}

         %\begin{array}{p{0.5\linewidth}l}
         \begin{tabular}{lccccc}
            \hline
            \noalign{\smallskip}
        Name  &   z & lg L$_{\rm 2-10keV}^a$ & lg L$_{\rm IR}$(tot)$^{a,b}$ &
    lg L$_{\rm UV-opt}$$^{a,c}$ & Refs.(X)$^d$ \\

           \noalign{\smallskip}
            \hline
            \noalign{\smallskip}

    PG0050+124  &   0.06 &  43.7  &  45.5 & 45.4  & 1  \\
    PG0054+144  &   0.17 &  44.3  &  45.5 & 45.7  & 1 \\
    PG0157+001  &   0.16 &  44.8  &  46.2 & 45.6  & 2 \\
    Q0710+45    &   0.05 &  44.3  &  45.0 & 45.2  & 3 \\
    PG0804+761  &    0.1 &  44.2  &  45.0 & 45.6  & 4 \\
    Q0844+34    &   0.06 &  43.3  &  44.7 & 45.2  & 5 \\
    PG1211+143  &   0.08 &  44.0  &  45.2 & 45.6  & 1 \\
    PG1307+085  &    0.1 &  44.1  &  44.7 & 45.5  & 6 \\
    IRAS13349+2438 & 0.1 &  44.4  &  45.8 & 45.5  & 1 \\
    PG1426+015  &   0.08 &  43.8  &  45.0 & 45.4  & 6 \\
    PG1440+356  &   0.07 &  43.8  &  45.0 & 45.3  & 1 \\
    PG1613+658  &   0.12 &  44.3  &  45.5 & 45.6  & 6 \\
    E1821+643   &   0.29 &  45.5  &  46.5 & 46.8  & 1 \\
    PG2130+099  &   0.06 &  43.5  &  45.0 & 45.3  & 6 \\

    \noalign{\smallskip}
            \hline
         \end{tabular}
\begin{flushleft}
$^a$ All luminosities are in units of erg~s$^{-1}$ \\
$^b$ Total (AGN+host) infrared luminosity in the range
8--1000$\mu$m estimated by using the IRAS data and
following the prescription in Sanders \& Mirabel (1996).\\
$^c$ Luminosity of the UV-optical blue bump of the AGN (see text).\\
$^d$ References for the X-ray data: 1--Reeves \& Turner (2000),
2--Boller et al. (2002), 3--Ceballos \& Barcons (1996), 4--George
et al. (2000), 5--Brinkmann et al. (2003), 6--Lawson \& Turner
(1997)
 \label{tab_qso}
\end{flushleft}
   \end{table*}

%\cite{reeves00},

 \label{lastpage}


\begin{thebibliography}{99}

%\bibitem[Alexander et al.(2002)]{alexander02} Alexander, D.~M.,
%Aussel, H., Bauer, F.~E., Brandt, W.~N., Hornschemeier, A.~E., Vignali, C.,
%Garmire, G.~P., \& Schneider, D.~P.\ 2002, ApJL, 568, L85

\bibitem[\protect\citeauthoryear{Aitken \& Roche}{1984}]{aitken84} Aitken, D.~K.~\&
Roche, P.~F.\ 1984, MNRAS, 208, 751

\bibitem[\protect\citeauthoryear{Alexander et al.}{2004}]{alexander04} Alexander, D.~M.,
Bauer, F.~E., Chapman, S.~C., Smail, I., Blain, A.~W., Brandt,
W.~N., \& Ivison, R.~J.\ 2004, in ``Multiwavelength Mapping of
Galaxy Formation and Evolution'', astro-ph/0401129

\bibitem[\protect\citeauthoryear{Alexander et al.}{2003}]{alexander03} Alexander, D.~M.~et
al.\ 2003, AJ, 125, 383

%\bibitem[Alonso-Herrero, Ward, \& Kotilainen(1996)]{alonso96}
%Alonso-Herrero, A., Ward, M.~J., \& Kotilainen, J.~K.\ 1996, MNRAS, 278,
%902

\bibitem[\protect\citeauthoryear{Alonso-Herrero et al.}{1998}]{alonso98} Alonso-Herrero, A., Simpson, C., Ward,
M.~J., \& Wilson, A.~S.\ 1998, ApJ, 495, 196

\bibitem[\protect\citeauthoryear{Alonso-Herrero et al.}{2001}]{alonso01} Alonso-Herrero,
A., Quillen, A.~C., Simpson, C., Efstathiou, A., \& Ward, M.~J.\ 2001, AJ,
121, 1369

\bibitem[\protect\citeauthoryear{Alonso-Herrero et al.}{2003}]{alonso03} Alonso-Herrero,
A., Quillen, A.~C., Rieke, G.~H., Ivanov, V.~D., \& Efstathiou, A.\ 2003,
AJ, 126, 81

\bibitem[\protect\citeauthoryear{Altieri et al.}{1999}]{altieri99} Altieri, B.~et al.\
1999, A\&A, 343, L65

\bibitem[\protect\citeauthoryear{Andreani et al.}{2003}]{andreani03a} Andreani, P.,
Cristiani, S., Grazian, A., La Franca, F., \& Goldschmidt, P.\ 2003, AJ,
125, 444

\bibitem[\protect\citeauthoryear{Andreani, Spinoglio, \& Malkan}{2003}]{andreani03b}
Andreani, P., Spinoglio, L., \& Malkan, M.~A.\ 2003, ApJ, 597, 759

\bibitem[\protect\citeauthoryear{Andreani, Franceschini, \& Granato}{1999}]{andreani99}
Andreani, P., Franceschini, A., \& Granato, G.\ 1999, MNRAS, 306, 161

\bibitem[\protect\citeauthoryear{Barger et al.}{2001}]{barger01} Barger, A.~J., Cowie,
L.~L., Steffen, A.~T., Hornschemeier, A.~E., Brandt, W.~N., \& Garmire,
G.~P.\ 2001, ApJL, 560, L23

\bibitem[\protect\citeauthoryear{Barger et al.}{2003}]{barger03} Barger, A.~J.~et al.\
2003, AJ, 126, 632

\bibitem[\protect\citeauthoryear{Bassani et al.}{1999}]{bassani99} Bassani, L., Dadina,
M., Maiolino, R., Salvati, M., Risaliti, G., della Ceca, R., Matt, G., \&
Zamorani, G.\ 1999, ApJS, 121, 473

\bibitem[\protect\citeauthoryear{Bertin et al.}{1997}]{bertin97} Bertin, E., Dennefeld, M., \& Moshir, M. 1997, A\&A,
323, 685

\bibitem[\protect\citeauthoryear{Brinkmann et al.}{2003}]{brinkman03} Brinkmann, W., Grupe, D.,
Branduardi-Raymont, G., \& Ferrero, E.\ 2003, A\&A, 398, 81

\bibitem[\protect\citeauthoryear{Boisson \& Durret}{1986}]{boisson86} Boisson, C.~\&
Durret, F.\ 1986, A\&A, 168, 32

\bibitem[\protect\citeauthoryear{Boller et al.}{2002}]{boller02} Boller,
T., Gallo, L.~C., Lutz, D., \& Sturm, E.\ 2002, MNRAS, 336, 1143

\bibitem[\protect\citeauthoryear{Boyle et al.}{2000}]{boyle00} Boyle, B.~J., Shanks, T.,
Croom, S.~M., Smith, R.~J., Miller, L., Loaring, N., \& Heymans, C.\ 2000,
MNRAS, 317, 1014

\bibitem[\protect\citeauthoryear{Brandt et al.}{2001}]{brandt01} Brandt, W.~N.~et al.\
2001, AJ, 122, 1

\bibitem[\protect\citeauthoryear{Ceballos \& Barcons}{1996}]{ceballos96} Ceballos, M.~T.~\&
Barcons, X.\ 1996, MNRAS, 282, 493

\bibitem[\protect\citeauthoryear{Cimatti et al.}{2002}]{cimatti02} Cimatti, A.~et al.\
2002, A\&A, 392, 395

\bibitem[\protect\citeauthoryear{Comastri et al.}{1995}]{comastri95}
Comastri, A., Setti, G., Zamorani, G., \& Hasinger, G.\ 1995, A\&A, 296, 1

\bibitem[\protect\citeauthoryear{Croom et al.}{2002}]{croom02} Croom, S.~M.~et al.\
2002, MNRAS, 337, 275

\bibitem[\protect\citeauthoryear{Della Ceca et al.}{2002}]{dellaceca02} Della Ceca, R.~et
al.\ 2002, ApJL, 581, L9

\bibitem[\protect\citeauthoryear{Devereux}{1989}]{devereux89} Devereux, N.~A.\ 1989, ApJ,
346, 126

\bibitem[\protect\citeauthoryear{Dole et al.}{2001}]{dole01} Dole, H., Gispert, R., Lagache, G., et al. 2001,
A\&A, 372, 702

\bibitem[\protect\citeauthoryear{Edelson, Malkan, \& Rieke}{1987}]{edelson87} Edelson,
R.~A., Malkan, M.~A., \& Rieke, G.~H.\ 1987, ApJ, 321, 233

\bibitem[\protect\citeauthoryear{Efstathiou \& Rowan-Robinson}{1995}]{efsta95}
Efstathiou, A.~\& Rowan-Robinson, M.\ 1995, MNRAS, 273, 649

\bibitem[\protect\citeauthoryear{Efstathiou et al.}{2000}]{efstathiou00} Efstathiou, A., Oliver, S., Rowan-Robinson, M., et al. 2000,
MNRAS, 319, 1169

\bibitem[\protect\citeauthoryear{Elbaz et al.}{1999}]{elbaz99} Elbaz, D.~et al.\ 1999,
A\&A, 351, L37

\bibitem[\protect\citeauthoryear{Elvis et al.}{1994}]{elvis94} Elvis, M.~et al.\ 1994,
ApJS, 95, 1

\bibitem[\protect\citeauthoryear{Fadda et al.}{2002}]{fadda02} Fadda, D., Flores, H.,
Hasinger, G., Franceschini, A., Altieri, B., Cesarsky, C.~J., Elbaz, D., \&
Ferrando, P.\ 2002, A\&A, 383, 838

\bibitem[\protect\citeauthoryear{Ferrarese \& Merritt}{2000}]{ferrarese00} Ferrarese, L.~\&
Merritt, D.\ 2000, ApJL, 539, L9

\bibitem[\protect\citeauthoryear{Fiore et al.}{2003}]{fiore03} Fiore, F.~et al.\ 2003,
A\&A, 409, 79

\bibitem[\protect\citeauthoryear{Frogel, Elias, \& Phillips}{1982}]{frogel82} Frogel,
J.~F., Elias, J.~H., \& Phillips, M.~M.\ 1982, ApJ, 260, 70

\bibitem[\protect\citeauthoryear{Fukazawa et al.}{2001}]{fuka01} Fukazawa, Y., Iyomoto,
N., Kubota, A., Matsumoto, Y., \& Makishima, K.\ 2001, A\&A, 374, 73

\bibitem[\protect\citeauthoryear{Gaskell et al.}{2003}]{gaskell03} Gaskell, C.~M., Goosmann, R.~W.,
Antonucci, R.~R.~J., \& Whysong, D.~H.\ 2003, ApJ submitted,
astro-ph/0309595

\bibitem[\protect\citeauthoryear{George et al.}{2000}]{george00} George, I.~M., Turner,
T.~J., Yaqoob, T., Netzer, H., Laor, A., Mushotzky, R.~F., Nandra, K., \&
Takahashi, T.\ 2000, ApJ, 531, 52

\bibitem[\protect\citeauthoryear{Giacconi et al.}{2001}]{giacconi01} Giacconi, R.~et al.\
2001, ApJ, 551, 624

\bibitem[\protect\citeauthoryear{Gilli, Salvati, \& Hasinger}{2001}]{gilli01} Gilli, R.,
Salvati, M., \& Hasinger, G.\ 2001, A\&A, 366, 407

\bibitem[\protect\citeauthoryear{Glass \& Moorwood}{1985}]{glass85} Glass, I.~S.~\&
Moorwood, A.~F.~M.\ 1985, MNRAS, 214, 429

\bibitem[\protect\citeauthoryear{Glass, Moorwood, \& Eichendorf}{1982}]{glass82} Glass,
I.~S., Moorwood, A.~F.~M., \& Eichendorf, W.\ 1982, A\&A, 107, 276

\bibitem[\protect\citeauthoryear{Glass}{2004}]{glass04} Glass,
I.~S., 2004, MNRAS, in press (astro-ph/0402289)

\bibitem[\protect\citeauthoryear{Granato \& Danese }{1994}]{granato94}
Granato, G.~L. \& Danese, L.\ 1994, MNRAS, 268, 235

\bibitem[\protect\citeauthoryear{Granato, Danese, \& Franceschini}{1997}]{granato97}
Granato, G.~L., Danese, L., \& Franceschini, A.\ 1997, ApJ, 486, 147

\bibitem[\protect\citeauthoryear{Granato et al.}{2004}]{granato04} Granato, G.~L., De
Zotti, G., Silva, L., Bressan, A., \& Danese, L.\ 2004, ApJ, 600, 580

\bibitem[\protect\citeauthoryear{Gruppioni et al.}{2002}]{gruppioni02} Gruppioni, C., Lari,
C., Pozzi, F., Zamorani, G., Franceschini, A., Oliver, S., Rowan-Robinson,
M., \& Serjeant, S.\ 2002, MNRAS, 335, 831

\bibitem[\protect\citeauthoryear{Guainazzi et al.}{2000}]{guainazzi00} Guainazzi, M., Matt,
G., Brandt, W.~N., Antonelli, L.~A., Barr, P., \& Bassani, L.\ 2000, A\&A,
356, 463

\bibitem[\protect\citeauthoryear{Haas et al.}{2003}]{haas03} Haas, M.~et al.\ 2003,
A\&A, 402, 87

\bibitem[\protect\citeauthoryear{Haas et al.}{2000}]{haas01} Haas, M., M{\" u}ller,
S.~A.~H., Chini, R., Meisenheimer, K., Klaas, U., Lemke, D., Kreysa, E., \&
Camenzind, M.\ 2000, A\&A, 354, 453

\bibitem[\protect\citeauthoryear{Hasinger}{2003}]{hasinger03} Hasinger, G.\ 2003, in
``The restless high energy universe'', astro-ph/0310804

\bibitem[\protect\citeauthoryear{Hasinger et al.}{2001}]{hasinger01} Hasinger, G.~et al.\
2001, A\&A, 365, L45

\bibitem[\protect\citeauthoryear{Hauser \& Dwek}{2001}]{hauser01} Hauser, M.~G.~\& Dwek,
E.\ 2001, ARA\&A, 39, 249

\bibitem[\protect\citeauthoryear{Iyomoto et al.}{2001}]{iyomoto01}
Iyomoto, N., Fukazawa, Y., Nakai, N., \& Ishihara, Y.\ 2001, ApJL, 561,
L69

\bibitem[\protect\citeauthoryear{Jaffe et al.}{2004}]{jaffe04}
Jaffe, W., et al. 2004, Nat, 429, 47

\bibitem[\protect\citeauthoryear{Kauffmann et al.}{2003}]{kauffmann03} Kauffmann, G.~et al.\
2003, MNRAS, 346, 1055

\bibitem[\protect\citeauthoryear{Kochanek et al.}{2001}]{kochanek01} Kochanek, C.~S.~et
al.\ 2001, ApJ, 560, 566

\bibitem[\protect\citeauthoryear{Kotilainen et al.}{1992}]{kotilainen92} Kotilainen, J.~K.,
Ward, M.~J., Boisson, C., Depoy, D.~L., \& Smith, M.~G.\ 1992, MNRAS, 256,
149

\bibitem[\protect\citeauthoryear{Krabbe, B{\" o}ker, \& Maiolino}{2001}]{krabbe01}
Krabbe, A., B{\" o}ker, T., \& Maiolino, R.\ 2001, ApJ, 557, 626

\bibitem[\protect\citeauthoryear{Kuraszkiewicz et al.}{2003}]{kura03} Kuraszkiewicz,
J.~K.~et al.\ 2003, ApJ, 590, 128

\bibitem[\protect\citeauthoryear{Imanishi \& Ueno}{1999}]{imanishi99} Imanishi, M.~\& Ueno,
S.\ 1999, ApJ, 527, 709

\bibitem[\protect\citeauthoryear{Ivanov et al.}{2000}]{ivanov00} Ivanov, V.~D., Rieke,
G.~H., Groppi, C.~E., Alonso-Herrero, A., Rieke, M.~J., \& Engelbracht,
C.~W.\ 2000, ApJ, 545, 190

\bibitem[\protect\citeauthoryear{La Franca et al.}{2004}]{lafranca04} La Franca, F.~et al.\
2004, AJ, in press (astro-ph/0403211)

\bibitem[\protect\citeauthoryear{Lawson \& Turner}{1997}]{lawson97} Lawson, A.~J.~\&
Turner, M.~J.~L.\ 1997, MNRAS, 288, 920

\bibitem[]{} Lonsdale, C. J., Hacking, P. B., Conrow, T. P., Rowan-Robinson,
M. 1990, ApJ, 358, 60

\bibitem[\protect\citeauthoryear{Malizia et al.}{1997}]{malizia97} Malizia, A., Bassani,
L., Stephen, J.~B., Malaguti, G., \& Palumbo, G.~G.~C.\ 1997, ApJS, 113,
311

\bibitem[\protect\citeauthoryear{Malkan \& Stecker}{2001}]{malkan01} Malkan, M.~A.~\&
Stecker, F.~W.\ 2001, ApJ, 555, 64

\bibitem[\protect\citeauthoryear{Mainieri et al.}{2002}]{mainieri02} Mainieri, V.,
Bergeron, J., Hasinger, G., Lehmann, I., Rosati, P., Schmidt, M., Szokoly,
G., \& Della Ceca, R.\ 2002, A\&A, 393, 425

\bibitem[\protect\citeauthoryear{Maiolino \& Rieke}{1995}]{maiolino95a} Maiolino, R.~\&
Rieke, G.~H.\ 1995, ApJ, 454, 95

\bibitem[\protect\citeauthoryear{Maiolino et al.}{1995}]{maiolino95b}
Maiolino, R., Ruiz, M., Rieke, G.~H., \& Keller, L.~D.\ 1995, ApJ, 446,
561

\bibitem[\protect\citeauthoryear{Maiolino et al.}{1998a}]{maiolino98a}
Maiolino, R., Krabbe, A., Thatte, N., \& Genzel, R.\ 1998, ApJ, 493, 650

\bibitem[\protect\citeauthoryear{Maiolino et al.}{1998b}]{maiolino98b} Maiolino, R., Salvati,
M., Bassani, L., Dadina, M., della Ceca, R., Matt, G., Risaliti, G., \&
Zamorani, G.\ 1998, A\&A, 338, 781

\bibitem[\protect\citeauthoryear{Maiolino, Marconi, \& Oliva}{2001}]{maiolino01a} Maiolino,
R., Marconi, A., \& Oliva, E.\ 2001, A\&A, 365, 37

\bibitem[\protect\citeauthoryear{Maiolino et al.}{2001}]{maiolino01b} Maiolino, R., Marconi,
A., Salvati, M., Risaliti, G., Severgnini, P., Oliva, E., La Franca, F., \&
Vanzi, L.\ 2001, A\&A, 365, 28

\bibitem[\protect\citeauthoryear{Maiolino}{2002}]{maiolino02} Maiolino, R.\ 2002, Reviews
of Modern Astronomy, 15, 179

\bibitem[\protect\citeauthoryear{Maiolino et al.}{2003}]{maiolino03} Maiolino, R.~et al.\
2003, MNRAS, 344, L59

\bibitem[\protect\citeauthoryear{Marconi et al.}{2000}]{marconi00} Marconi, A., Oliva, E.,
van der Werf, P.~P., Maiolino, R., Schreier, E.~J., Macchetto, F., \&
Moorwood, A.~F.~M.\ 2000, A\&A, 357, 24

\bibitem[\protect\citeauthoryear{Marconi \& Hunt}{2003}]{marconi03} Marconi, A.~\& Hunt,
L.~K.\ 2003, ApJL, 589, L21

\bibitem[\protect\citeauthoryear{Matt et al.}{1997}]{matt97} Matt, G.~et al.\ 1997,
A\&A, 325, L13

\bibitem[\protect\citeauthoryear{Matt et al.}{1999}]{matt99} Matt, G.~et al.\ 1999,
A\&A, 341, L39

\bibitem[\protect\citeauthoryear{Matute et al.}{2002}]{matute02} Matute, I.~et al.\ 2002,
MNRAS, 332, L11

\bibitem[\protect\citeauthoryear{Mazzei et
al.}{2001}]{2001NewA....6..265M} Mazzei, P., Aussel, H., Xu, C.,
Salvo, M., De Zotti, G., \& Franceschini, A. 2001, NewA, 6, 265

\bibitem[\protect\citeauthoryear{McAlary et al.}{1983}]{mcalary83}
McAlary, C.~W., McLaren, R.~A., McGonegal, R.~J., \& Maza, J.\ 1983, ApJS,
52, 341

\bibitem[\protect\citeauthoryear{McLeod \& Rieke}{1994a}]{mcleod94a} McLeod, K.~K.~\&
Rieke, G.~H.\ 1994, ApJ, 431, 137

\bibitem[\protect\citeauthoryear{McLeod \& Rieke}{1994b}]{mcleod94b} McLeod, K.~K.~\&
Rieke, G.~H.\ 1994, ApJ, 420, 58

\bibitem[\protect\citeauthoryear{Minezaki et al.}{2004}]{minezaki04} Minezaki, T., Yoshii,
Y., Kobayashi, Y., Enya, K., Suganuma, M., Tomita, H., Aoki, T., \&
Peterson, B.~A.\ 2004, ApJ, 600, L35

\bibitem[\protect\citeauthoryear{Moustakas et al.}{1997}]{moustakas97} Moustakas, L.~A.,
Davis, M., Graham, J.~R., Silk, J., Peterson, B.~A., \& Yoshii, Y.\ 1997,
ApJ, 475, 445

\bibitem[\protect\citeauthoryear{Nenkova, Ivezi{\' c}, \& Elitzur}{2002}]{nenkova02}
Nenkova, M., Ivezi{\' c}, {\v Z}., \& Elitzur, M.\ 2002, ApJ, 570, L9

\bibitem[\protect\citeauthoryear{Oliva et al.}{1999}]{oliva99}
Oliva, E., Origlia, L., Maiolino, R., \& Moorwood, A.~F.~M.\ 1999, A\&A,
350, 9

\bibitem[\protect\citeauthoryear{Oliver et al.}{1997}]{oliver97} Oliver, S.~J.~et al.\
1997, MNRAS, 289, 471

\bibitem[\protect\citeauthoryear{Oliver et al.}{2002}]{oliver02} Oliver, S.~et al.\ 2002,
MNRAS, 332, 536

\bibitem[\protect\citeauthoryear{Pearson \& Rowan-Robinson}{1996}]{pearson96} Pearson, C., \& Rowan-Robinson, M. 1996, MNRAS, 283,
174

\bibitem[\protect\citeauthoryear{Perola et al.}{2002}]{perola02} Perola, G.~C., Matt, G.,
Cappi, M., Fiore, F., Guainazzi, M., Maraschi, L., Petrucci, P.~O., \&
Piro, L.\ 2002, A\&A, 389, 802

\bibitem[\protect\citeauthoryear{Papadopoulos \& Seaquist}{1999}]{papa99} Papadopoulos,
P.~P.~\& Seaquist, E.~R.\ 1999, ApJL, 514, L95

\bibitem[\protect\citeauthoryear{Pier \& Krolik}{1993}]{pier93} Pier, E.~A.~\& Krolik,
J.~H.\ 1993, ApJ, 418, 673

\bibitem[\protect\citeauthoryear{Quillen et al.}{2001}]{quillen01} Quillen, A.~C.,
McDonald, C., Alonso-Herrero, A., Lee, A., Shaked, S., Rieke, M.~J., \&
Rieke, G.~H.\ 2001, ApJ, 547, 129

\bibitem[\protect\citeauthoryear{Reeves \& Turner}{2000}]{reeves00} Reeves, J.~N.~\&
Turner, M.~J.~L.\ 2000, MNRAS, 316, 234

\bibitem[\protect\citeauthoryear{Reynolds}{1997}]{reynolds97} Reynolds, C.~S.\ 1997,
MNRAS, 286, 513

\bibitem[\protect\citeauthoryear{Rieke}{1978}]{rieke78} Rieke, G.~H.\ 1978, ApJ, 226,
550

\bibitem[\protect\citeauthoryear{Risaliti, Maiolino, \& Salvati}{1999}]{risaliti99}
Risaliti, G., Maiolino, R., \& Salvati, M.\ 1999, ApJ, 522, 157

\bibitem[\protect\citeauthoryear{Risaliti, Elvis, \& Gilli}{2002}]{risaliti02} Risaliti,
G., Elvis, M., \& Gilli, R.\ 2002, ApJL, 566, L67

\bibitem[\protect\citeauthoryear{Rodichiero et al.}{2003}]{rodighiero03} Rodighiero, G., Lari, C., Franceschini, A., Gregnanin, A., Fadda,
D. 2003, MNRAS, 343, 1155

\bibitem[\protect\citeauthoryear{Rudy, Levan, \& Rodriguez-Espinosa}{1982}]{rudy82}
Rudy, R.~J., Levan, P.~D., \& Rodriguez-Espinosa, J.~M.\ 1982, AJ, 87, 598

\bibitem[\protect\citeauthoryear{Sanders \& Mirabel}{1996}]{sanders96} Sanders, D.~B.~\&
Mirabel, I.~F.\ 1996, ARA\&A, 34, 749

\bibitem[\protect\citeauthoryear{Saracco et al.}{2001}]{saracco01} Saracco, P., Giallongo,
E., Cristiani, S., D'Odorico, S., Fontana, A., Iovino, A., Poli, F., \&
Vanzella, E.\ 2001, A\&A, 375, 1

\bibitem[\protect\citeauthoryear{Sato et al.}{2003}]{sato03} Sato, Y.~et al.\ 2003,
A\&A, 405, 833

\bibitem[\protect\citeauthoryear{Schinnerer et al.}{2000}]{eva00} Schinnerer, E.,
Eckart, A., Tacconi, L.~J., Genzel, R., \& Downes, D.\ 2000, ApJ, 533, 850

\bibitem[\protect\citeauthoryear{Scoville, Becklin, Young, \& Capps}{1983}]{scoville83}
Scoville, N.~Z., Becklin, E.~E., Young, J.~S., \& Capps, R.~W.\ 1983, ApJ,
271, 512

\bibitem[\protect\citeauthoryear{Serjeant et al.}{2000}]{serjeant00} Serjeant, S.~et al.\
2000, MNRAS, 316, 768

\bibitem[\protect\citeauthoryear{Setti \& Woltjer}{1989}]{setti89} Setti, G.~\& Woltjer,
L.\ 1989, A\&A, 224, L21

\bibitem[\protect\citeauthoryear{Severgnini et al.}{2000}]{severgnini00} Severgnini, P.~et
al.\ 2000, A\&A, 360, 457

\bibitem[\protect\citeauthoryear{Siebenmorgen et al.}{1997}]{siebenmorgen97} Siebenmorgen, R., Moorwood, A.,
Freudling, W., \& Kaeufl, H.~U.\ 1997, A\&A, 325, 450

\bibitem[\protect\citeauthoryear{Silva et al.}{1998}]{silva98}
Silva, L., Granato, G.~L., Bressan, A., \& Danese, L.\ 1998, ApJ, 509, 103

\bibitem[\protect\citeauthoryear{Silva et al.}{2004}]{silva04}
Silva, L., De~Zotti, G., Granato, G.~L., Maiolino, R., \& Danese,
L.\ 2004, submitted (astro-ph/0403166)

\bibitem[\protect\citeauthoryear{Simpson}{1998}]{simpson98} Simpson, C.\ 1998, ApJ, 509,
653

\bibitem[\protect\citeauthoryear{Sparks et al.}{1986}]{sparks86} Sparks,
W.~B., Hough, J.~H., Axon, D.~J., \& Bailey, J.\ 1986, MNRAS, 218, 429

\bibitem[\protect\citeauthoryear{Spinoglio, Andreani, \& Malkan}{2002}]{spinoglio02}
Spinoglio, L., Andreani, P., \& Malkan, M.~A.\ 2002, ApJ, 572, 105

\bibitem[\protect\citeauthoryear{Surace, Sanders, \& Evans}{2001}]{surace01} Surace,
J.~A., Sanders, D.~B., \& Evans, A.~S.\ 2001, AJ, 122, 2791

\bibitem[\protect\citeauthoryear{Taniguchi et al.}{1997}]{taniguchi97} Taniguchi, Y.~et al.\
1997, A\&A, 328, L9

\bibitem[\protect\citeauthoryear{Telfer et al.}{2002}]{telfer02}
Telfer, R.~C., Zheng, W., Kriss, G.~A., \& Davidsen, A.~F.\ 2002, ApJ,
565, 773

\bibitem[\protect\citeauthoryear{Tresch-Fienberg et al.}{1987}]{tresch87}
Tresch-Fienberg, R., Fazio, G.~G., Gezari, D.~Y., Lamb, G.~M., Shu, P.~K.,
Hoffmann, W.~F., \& McCreight, C.~R.\ 1987, ApJ, 312, 542

\bibitem[\protect\citeauthoryear{Totani et al.}{2001}]{totani01} Totani, T., Yoshii, Y.,
Maihara, T., Iwamuro, F., \& Motohara, K.\ 2001, ApJ, 559, 592

\bibitem[\protect\citeauthoryear{Turner et al.}{1999}]{turner99}
Turner, T.~J., George, I.~M., Nandra, K., \& Turcan, D.\ 1999, ApJ, 524,
667

\bibitem[\protect\citeauthoryear{Ueda et al.}{2003}]{ueda03} Ueda,
Y., Akiyama, M., Ohta, K., \& Miyaji, T.\ 2003, ApJ, 598, 886

\bibitem[\protect\citeauthoryear{Vanden Berk et al.}{2001}]{berk01} Vanden Berk,
D.~E.~et al.\ 2001, AJ, 122, 549

\bibitem[\protect\citeauthoryear{Vignali \& Comastri}{2002}]{vignali02} Vignali, C.~\&
Comastri, A.\ 2002, A\&A, 381, 834

\bibitem[\protect\citeauthoryear{Ward et al.}{1982}]{ward82} Ward, M., Allen, D.~A.,
Wilson, A.~S., Smith, M.~G., \& Wright, A.~E.\ 1982, MNRAS, 199, 953

\bibitem[\protect\citeauthoryear{Ward et al.}{1987}]{ward87} Ward, M., Elvis, M.,
Fabbiano, G., Carleton, N.~P., Willner, S.~P., \& Lawrence, A.\ 1987, ApJ,
315, 74

\bibitem[\protect\citeauthoryear{Wright et al.}{1988}]{wright88} Wright, G.~S., Joseph,
R.~D., Robertson, N.~A., James, P.~A., \& Meikle, W.~P.~S.\ 1988, MNRAS,
233, 1

\bibitem[\protect\citeauthoryear{Zitelli et al.}{1993}]{zitelli93} Zitelli, V., Granato,
G.~L., Mandolesi, N., Wade, R., \& Danese, L.\ 1993, ApJS, 84, 185

\end{thebibliography}
\end{document}